\documentclass[article,traditabstract]{aa} 
\usepackage{natbib}
\usepackage{graphicx}
\usepackage{txfonts}
\usepackage{enumitem}

\newcommand{\kms}{km~s\ensuremath{^{-1}}}
\newcommand{\msun}{M$_{\odot}$}
\newcommand{\mstar}{M$_{\ast}$}

\newcommand{\mh}{$M_{{\rm H_2}}$}
\newcommand{\mhi}{$M_{{\rm HI}}$}
\newcommand{\xco}{$\alpha_{{\rm CO}}$}
\newcommand{\gdr}{$\delta_{{\rm\scriptsize{GDR}}}$}
\newcommand{\gdrm}{$\delta_{{\rm\scriptsize{GDR}}}(\mu_0)$}
\newcommand{\xcounits}{\msun~(K \kms\ pc$^{2}$)$^{-1}$}

\begin{document} 
\title{Dust temperature and CO$\rightarrow$H$_2$ conversion factor variations \\
in the SFR-$M_{\ast}$ plane\thanks{Based on observations carried out with the IRAM Plateau de Bure Interferometer. IRAM is supported by INSU/CNRS (France), MPG (Germany) and IGN (Spain). Based also on observations carried out by the \textit{Herschel} space observatory. \textit{Herschel} is an ESA space observatory with science instruments provided by European-led Principal Investigator consortia and with important participation from NASA.}}
\author{B.~Magnelli\inst{1}
        \and
        A.~Saintonge\inst{1}
        \and
        D.~Lutz\inst{1}
        \and
        L.\,J.~Tacconi\inst{1}
        \and
        S.~Berta\inst{1}
        \and
        F.~Bournaud\inst{2}
        \and
        V.~Charmandaris\inst{3,4,5}
        \and
        H.~Dannerbauer\inst{6}
        \and
        D.~Elbaz\inst{2}
        \and
        N.\,M.~F\"orster-Schreiber\inst{1}
        \and
        J.~Graci\'a-Carpio\inst{1}
        \and
        R.~Ivison\inst{7}
        \and
        R.~Maiolino\inst{8}
        \and
        R.~Nordon\inst{1}
        \and
	P.~Popesso\inst{1}
        \and
	G.~Rodighiero\inst{9}
	\and
	P. Santini\inst{10}
	\and
	S.~Wuyts\inst{1}
              }
\offprints{B. Magnelli \& A. Saintonge, \email{magnelli@mpe.mpg.de, amelie@mpe.mpg.de}}
\institute{
Max-Planck-Institut f\"{u}r Extraterrestrische Physik (MPE), Postfach 1312, 85741 Garching, Germany
\and
Laboratoire AIM, CEA/DSM-CNRS-Universit{\'e} Paris Diderot, IRFU/Service
d'Astrophysique,
B\^at.709, CEA-Saclay, 91191 Gif-sur-Yvette Cedex, France.
\and
Department of Physics and Institute of Theoretical and Computational Physics, University of Crete, GR-71002, Heraklion, Greece
\and
Chercheur Associ\'e, Observatoire de Paris, F-75014 Paris, France
\and
IESL/Foundation for Research and Technology-Hellas, P.O. Box 1527, 71110 Heraklion, Greece
\and
Universit\"{a}t Wien, Institut f\"{u}r Astronomie, T\"{u}rkenschanzstra\ss e 17, 1180 Wien, \"{O}sterreich
\and
UK Astronomy Technology Centre, Royal Observatory, Blackford Hill, Edinburgh EH9 3HJ, UK
\and
Cavendish Laboratory, University of Cambridge, 19 J.J. Thomson Ave., Cambridge CB3 0HE, UK
\and
Dipartimento di Astronomia, Universita di Padova, Vicolo dell'Osservatorio 3, 35122 Padova, Italy
\and
INAF-Osservatorio Astronomico di Roma, via di Frascati 33, 00040 Monte Porzio Catone, Italy
}
\date{Received ??; accepted ??}
\abstract{
Deep {\it Herschel} PACS/SPIRE imaging and $^{12}$CO(2-1) line luminosities from the IRAM Plateau de Bure Interferometer are combined for a sample of 17 galaxies at $z>1$ from the GOODS-N field.  The sample includes galaxies both on and above the main sequence (MS) traced by star-forming galaxies in the SFR-\mstar\ plane.  The far-infrared data are used to derive dust masses, $M_{{\rm dust}}$, following the Draine \& Li (2007) models.  Combined with an empirical prescription for the dependence of the gas-to-dust ratio on metallicity (\gdrm), the CO luminosities and $M_{{\rm dust}}$ values are used to derive for each galaxy the CO-to-H$_2$ conversion factor, \xco.   Like in the local Universe, the value of \xco\ is a factor of $\sim$5 smaller in starbursts compared to normal star-forming galaxies (SFGs).  We additionally uncover a relation between \xco\ and dust temperature ($T_{{\rm dust}}$; \xco\ decreasing with increasing $T_{{\rm dust}}$) as obtained from modified blackbody fits to the far-infrared data.  While the absolute normalization of the \xco$(T_{{\rm dust}})$ relation is uncertain, the global trend is robust against possible systematic biases in the determination of $M_{{\rm dust}}$, \gdrm\ or metallicity.  Although we cannot formally distinguish between a step and a smooth evolution of \xco\ with the dust temperature, we can unambiguously conclude that in galaxies of near-solar metallicity, a critical value of $T_{{\rm dust}}=30$K can be used to determine whether the appropriate \xco\ is closer to the ``starburst" value (1.0 \msun (K \kms\ pc$^{2}$)$^{-1}$, when $T_{{\rm dust}}$$\,>30\,$K) or closer to the Galactic value (4.35 \msun (K \kms\ pc$^{2}$)$^{-1}$, when $T_{{\rm dust}}$$\,<30\,$K).  This indicator has the great advantage of being less subjective than visual morphological classifications of mergers/SFGs, which can be difficult at high $z$ because of the clumpy nature of SFGs. Using $T_{{\rm dust}}$ to select the appropriate \xco\ is also more indicative of ISM conditions than a fixed $L_{{\rm IR}}$ criterion.  In the absence of far-infrared data, the offset of a galaxy from the star formation main sequence (i.e., ${\rm \Delta log(SSFR)_{MS}=log[SSFR(galaxy)/SSFR_{MS}}(M_{\ast},z)]$) can be used to identify galaxies requiring the use of an \xco\ conversion factor lower than the Galactic value (i.e., when ${\rm \Delta log(SSFR)_{MS}} \gtrsim 0.3\,$dex). 
}
\authorrunning{Magnelli et al. }
\titlerunning{Molecular gas properties of galaxies in the SFR-$M_{\ast}$ plane}
\maketitle
\section{Introduction}

A correlation between the star formation rate (SFR) and the stellar mass (\mstar) of star-forming galaxies has been observed over the last $10\,$Gyr of lookback time \citep[SFR$\,\propto\,M_{\ast}^\alpha$, with $0.5<\alpha<1.0$; ][]{brinchmann_2004,schiminovich_2007,noeske_2007a,elbaz_2007,daddi_2007a,pannella_2009,rodighiero_2010b,oliver_2010,magdis_2010b,karim_2011,mancini_2011}.  The existence of this main sequence (MS) of star formation is usually interpreted as evidence that the bulk of the galaxy population is forming stars gradually with long duty cycles. In this interpretation, galaxies situated on the MS are consistent with a secular mode of star formation, likely sustained by a continuous gas accretion from the IGM and along the cosmic web \citep{dekel_2009,dave_2010}, while star-forming galaxies located far above the MS are strong starbursts with short duty-cycles, mainly triggered by major mergers \citep[see, e.g.][]{engel_2010}.
In the last years, independent observational tests have given weight to this interpretation.
For example, it is observed that galaxies situated on the MS of star-formation have a disk-like morphology, relatively low SFR surface density \citep{wuyts_2011b}, cold dust temperature \citep[][see also, Magnelli et al. in prep.]{magnelli_2012} and large polycyclic aromatic hydrocarbons (PAHs) equivalent width \citep[EW;][]{nordon_2012,elbaz_2011}.
On the other hand, off-MS galaxies (i.e., situated above the MS) have cuspier morphology, higher SFR surface density, hotter dust temperature and smaller PAHs EW \citep[][]{wuyts_2011b,magnelli_2012,nordon_2012,elbaz_2011}.

Off-MS galaxies reaching very high specific star formation rates (SSFR$=$SFR$/$\mstar) do so by having enhanced star formation efficiencies (SFR$/$$M_{{\rm H_2}}$; where $M_{{\rm H_2}}$ is the molecular gas mass), as observed in local ULIRGs and high redshift submillimetre galaxies \citep[e.g.][]{gao_2004,daddi_2010,genzel_2010,saintonge_2011b}.  These results are however complicated by the fact that the CO-to-H$_2$ conversion factor, \xco, also likely changes between normal star-forming discs on the main sequence and merger-driven starbursts \citep{solomon_1997,tacconi_2008}.  Having either a measurement of \xco\ in any given galaxy, or a reliable prescription to assign its value is therefore critical to study variations in star formation efficiency across the galaxy population. 

There are several methods to measure \xco\ in high-redshift galaxies. 
For example a careful analysis of the stellar and dynamical masses of luminous SMGs has been used to determine their CO-to-H$_2$ conversion factor, \xco\ \citep{tacconi_2008}.
Although successful, this method requires very high quality interferometric CO observations (i.e., spectroscopically and spatially resolved) in order to determine the dynamical masses of galaxies.
Consequently, before the advent in the near future of the full capabilities of the Atacama Large Millimeter Array (ALMA), this method will be only available for a handful of CO-bright galaxies.
In order to constrain \xco\ for a larger and fainter sample of galaxies another method is needed.
In the local Universe, a method used to derive \xco\ relies on the determination of the gas mass of a galaxy via the measurement of its dust mass, using far-infrared observations, and an assumed gas-to-dust ratio \citep[see, e.g.,][]{leroy_2011}.
Thanks to the advent of the \textit{Herschel} Space Observatory, this method can now be applied to a large sample of high-redshift galaxies.
\citet{magdis_2011b,magdis_2012} have pioneered this method on a sample including a dozen high-redshift galaxies. 
Based on this recent progress we apply this method to a larger sample of high-redshift galaxies.

Here we compile a large sample of 17 $z>1$ galaxies in the GOODS-N field with both deep {\it Herschel} far-infrared measurements and accurate IRAM Plateau de Bure Interferometer (PdBI) line luminosities.  After describing the sample selection, the data products, and the derived quantities in \S\S \ref{sample}-\ref{analysis},  we combine these measurements to measure the value of the conversion factor \xco\ for each galaxy, discuss its trends with different global galaxy properties, and address the impact of possible systematic biases (\S \ref{alphaco}).  Finally, the origin and impact of the observed relations between \xco\ and quantities such as $T_{{\rm dust}}$ and MS offset are discussed in \S \ref{sec: discussion} before the main conclusions are summarized in \S \ref{summary}.

All rest-frame quantities derived in this work assume a \citet{chabrier_2003} IMF, and a cosmology with $H_0=71$\kms\ Mpc$^{-1}$, $\Omega_m=0.27$ and $\Omega_{\Lambda}=0.73$.

\section{Sample Selection}
 \label{sample}
 
Our goal is to study the link between molecular gas, dust, star-formation activity and stellar mass in high-redshift galaxies.
Such a study requires using direct tracers of molecular gas and dust, and probing a large dynamic range in the SFR-\mstar\ plane.  One of the deep fields best suited for this study is GOODS-N, having a broad multi-wavelength coverage including some of the deepest {\it Herschel} maps (with both the PACS and SPIRE instrument, see Section \ref{subsec: herschel}), and with its northern location, being a privileged target for the IRAM Plateau de Bure Interferometer (PdBI).  The interferometric observations produce the CO line fluxes from which total molecular gas masses can be extrapolated, while the dust content of high-redshift galaxies can be studied using the wide far-infrared wavelength coverage from {\it Herschel}.  

\begin{figure*}
\center
\includegraphics[width=9.cm]{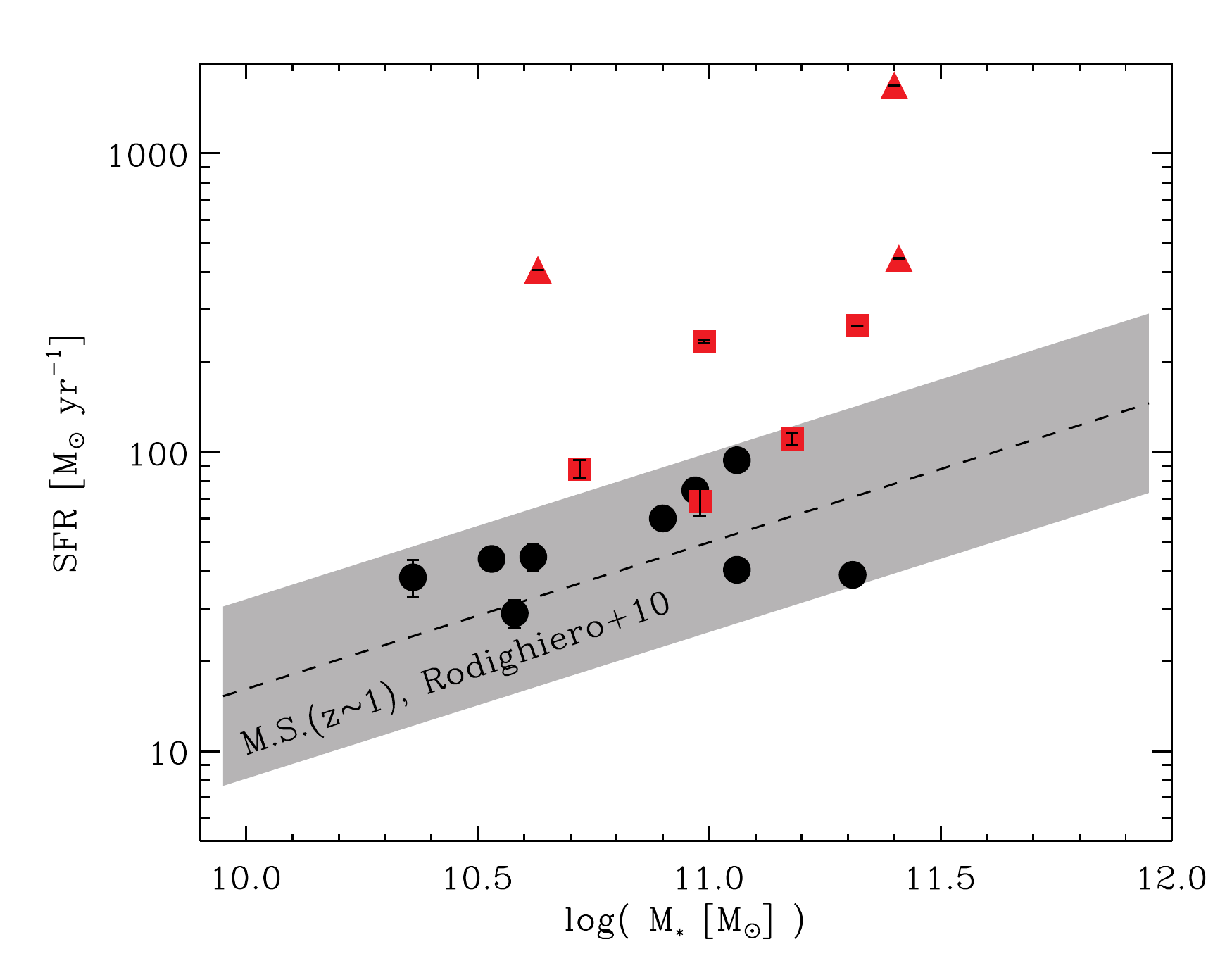}
\includegraphics[width=9.cm]{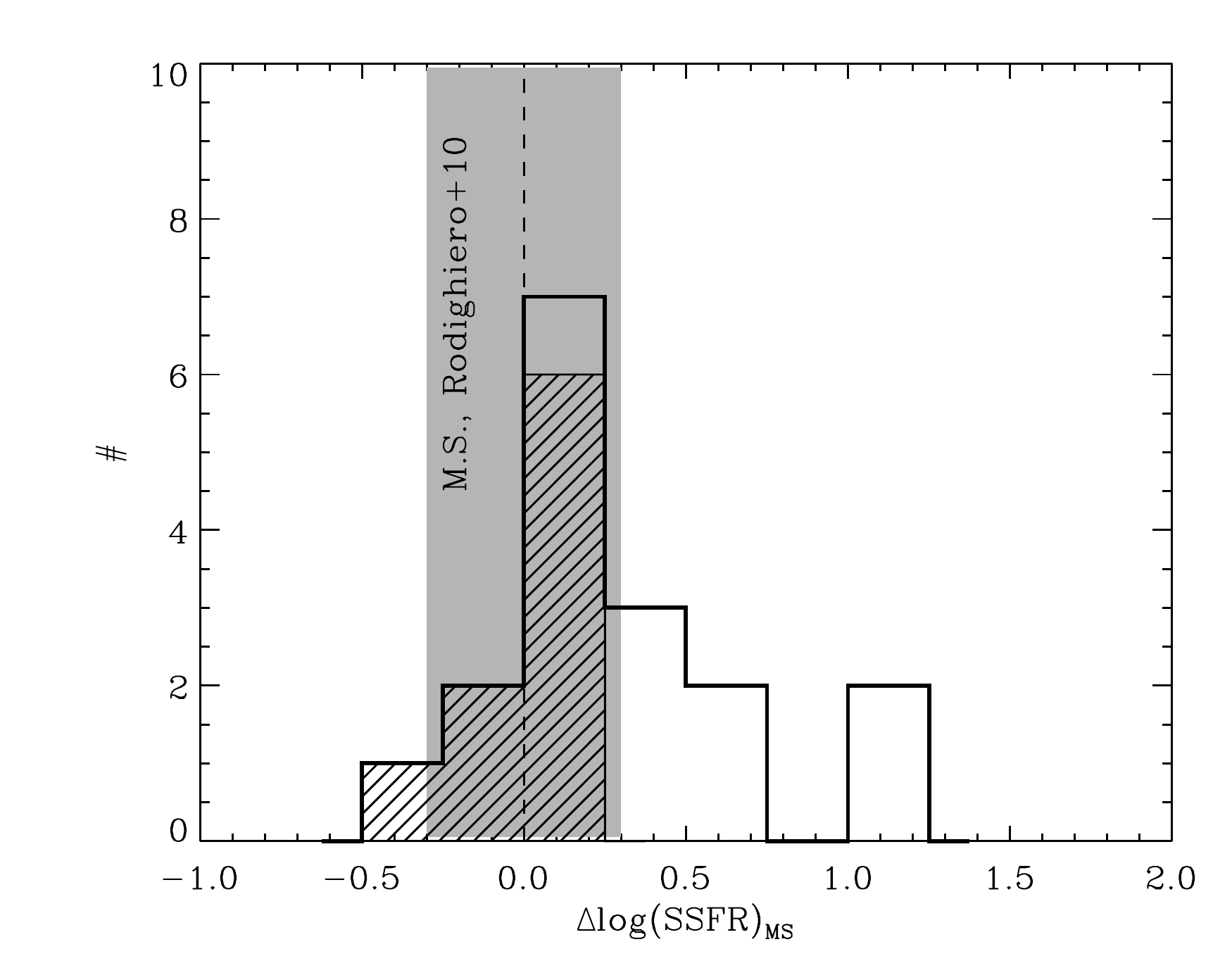}
\caption{ \label{Fig: SFRmstar}
(\textit{Left}) The $z\sim1$ star formation main sequence from \citet{rodighiero_2010b} and the position of the GOODS-N sample with CO measurements and PACS/SPIRE photometry.
Galaxies with new IRAM CO observations are shown as black circles while the red symbols are for galaxies with CO data from the literature \citep{frayer_2008,daddi_2010,carilli_2010,ivison_2011}.
Among the red symbols, triangles stand for galaxies referred to SMGs.
(\textit{Right}) Distribution of distance from the star formation main sequence ($\Delta$log$({\rm SSFR})_{MS}$), for the galaxies with the new IRAM CO observations (dashed histogram), and the combined sample including galaxies CO measurements from the literature (open histogram).
Bins of 0.25 dex in $\Delta$log$({\rm SSFR})_{MS}$ are used.
}
\end{figure*}

We found 9 GOODS-N galaxies in the published literature with CO(1-0) or CO(2-1) measurements, HDF169 (GN26) in \citet{frayer_2008},  BzK-4171, BzK-21000, BzK-16000, BzK-17999, BzK-12591 and BzK-25536 in \citet{daddi_2010}, HDF242 (GN19) in \citet{ivison_2011}, and finally GN20 in \citet{carilli_2010}.   Since BzK-25536 is not detected in the deep PACS and SPIRE images (see Section \ref{subsec: herschel}), it is excluded from our study.  The remaining 8 galaxies\footnote{We note among these 8 galaxies, BzK-21000 and GN20 were initially analyzed in \citet{magdis_2011b}, while the remaining 6 are also presented in \citet{magdis_2012}.} are shown in the SFR-\mstar\ plane in the left panel of Figure \ref{Fig: SFRmstar} (see Section \ref{subsec: herschel} for a complete description of the data used to derive SFR and \mstar).
All of these galaxies lie above the mean star formation main sequence, with three SMGs being the strongest outliers. In order to study the evolution of the molecular gas content of galaxies as a function of their location in the SFR-\mstar\ plane, we have obtained new IRAM PdBI CO data for galaxies located on or even below this main sequence of star-formation.

At $z$$\,\thicksim\,$$1$, the \textit{Herschel} observations of the GOODS-N field allow for the detection of galaxies with SFR$\gtrsim10\,$M$_{\odot}$yr$^{-1}$, probing well into the main sequence for galaxies with \mstar$>10^{10.2}$\msun\ \citep[see, e.g.,][]{rodighiero_2010b,noeske_2007a,elbaz_2007,karim_2011}.  Here we use the definition of the main-sequence from \citet{rodighiero_2010b}, i.e., ${\rm log(SSFR)_{MS}=\alpha\,log}(M_{\ast})$$\,+\,$$\beta$ where $(\alpha\,,\,\beta)$$\,=\,$$(-0.27,2.6)$, $(-0.51,5.3)$ and $(-0.49,5.2)$ at $0.5<z<1.0$, $1.0<z<1.5$ and $z>1.5$, respectively.
This choice is motivated by the use of the FIR as a star-formation indicator both in \citet{rodighiero_2010b} and in this study as well as by the use of fully consistent stellar mass estimates\footnote{Results of \citet{rodighiero_2010b} are corrected for different IMF assumptions, namely from an Salpeter IMF in \citet{rodighiero_2010b} to a \citet{chabrier_2003} IMF in our study. Thus, SFRs and stellar masses from \citet{rodighiero_2010b} are divided by a factor $1.7$}.
However, none of our results strongly depend on this specific definition.

We selected 9 IRAM PdBI CO targets from the complete volume-limited sample of galaxies with $1.0<z<1.3$, SFR$\gtrsim15\,$M$_{\odot}$\,yr$^{-1}$ and \mstar$>10^{10.2}\,$\msun (see table \ref{tab:pdbi}).  In addition, we required the galaxies to have reliable spectroscopic redshift measurements from \citet{barger_2008}.  We note that our sample is not intended to be complete, but rather to sample as uniformly as possible the SFR-\mstar\ plane (see Fig. \ref{Fig: SFRmstar}).   The combined sample of 17 galaxies, including the objects extracted from the literature and our new CO targets, spans almost two orders of magnitude in SFR at fixed stellar mass; the final distribution of main sequence offset (i.e., ${\rm \Delta log(SSFR)_{MS}=log[SSFR(galaxy)/SSFR_{MS}}(M_{\ast},z)]$, where ${\rm SSFR(galaxy)}$ is the specific star-formation rate of our galaxies and ${\rm SSFR_{MS}(M_{\ast},z)}$ is the specific star-formation rate of MS galaxies at the redshifts and stellar masses of our galaxies) is shown in the right panel of Fig. \ref{Fig: SFRmstar}.

\begin{table*}
\scriptsize
\caption{\label{tab:pdbi} Measured and derived properties}
\centering
\begin{tabular}{ c cc  c c c c c c c c c} 
\hline \hline
Galaxy & \multicolumn{2}{c}{Position} & Redshift & Obs. CO & $S_{{\rm CO}}\Delta V$ & $W_{{\rm CO}}$ & Flag$^{{\rm a}}$ & $L^{\prime}_{CO(1-0)}$ & Ref. CO & log(\mstar) & Metallicity$^{{\rm b}}$\\
 & { RA} & {DEC} & & Transition & {Jy km s$^{-1}$} & { km s$^{-1}$} &  & { 10$^9\ $K$\,$km$\,$s$^{-1}$pc$^2$} & & { log(M$_{\odot}$)} & { 12+log(O/H)} \\
\hline
PEPJ123712+621753 & 12      37      12.15   &   62      17      53.95  & 1.249   & 2$-$1&  $0.15\pm0.04$ &  $409\pm170$ & 1 & $4.2\pm1.1$ &This work & 10.62 & 8.73    \\
PEPJ123709+621507 & 12      37      09.13   &   62      15      07.92  & 1.224   & 2$-$1&  $0.28\pm0.06$ &  $231\pm90$ & 2 & $7.5\pm1.6$ &This work & 10.36 & 8.67      \\
PEPJ123759+621732 & 12      37      59.47   &   62      17      32.86  & 1.084   & 2$-$1&  $0.44\pm0.11$ &  $406\pm115$ & 1 &$9.3\pm2.3$ &This work & 10.58 & 8.72       \\
PEPJ123721+621346 & 12      37      21.45   &   62      13      46.06  & 1.021   & 2$-$1&  $0.39\pm0.13$ &  $236\pm160$ & 2 & $7.3\pm2.4$ &This work & 10.53 & 8.71     \\
PEPJ123615+621008 & 12      36      15.19   &   62      10      08.65  & 1.027   & 2$-$1&  $<0.42$ &  $\dots$ & 3 & $7.9\pm1.5$ &This work & 10.90 & 8.78      \\
PEPJ123634+620627 & 12      36      34.50   &   62      06      27.94  & 1.215   & 2$-$1&  $<0.27$ &  $\dots$ & 3 & $7.1\pm1.3$ &This work & 11.06 & 8.80     \\
PEPJ123633+621005 & 12      36      33.65   &   62      10      05.77  & 1.016   & 2$-$1&  $0.78\pm0.19$ &  $393\pm120$ & 2 & $14.4\pm3.5$ &This work & 10.97 & 8.79    \\
PEPJ123646+621141 & 12      36      46.18   &   62      11      41.99  & 1.016   & 2$-$1&  $0.40\pm0.09$ &  $157\pm60$ & 1 & $7.4\pm1.6$ &This work & 11.06 & 8.80      \\ 
PEPJ123750+621600 & 12      37      50.89   &   62      16      00.69  & 1.170   & 2$-$1&  $0.53\pm0.08$ &  $649\pm110$ & 1 & $13.0\pm1.9$ &This work & 11.31 & 8.83    \\
HDF169        & 12      36      34.53   &   62      12      40.92  & 1.224  & 2$-$1  & $3.45\pm0.93$ &  560  & $\dots$ & $92.3\pm24.9$ & Frayer+\citeyear{frayer_2008} & 10.63  & 8.73 \\
BzK-4171     & 12      36      26.52   &   62       08      35.30  & 1.465  & 2$-$1  &  $0.65\pm0.08$ &  530 & $\dots$ & $24.6\pm3.0$ & Daddi+\citeyear{daddi_2010} & 10.72  &  8.75 \\
BzK-21000  & 12      37      10.64   &   62      22      34.17  & 1.523  & 2$-$1  & $0.64\pm0.07$ & 444   & $\dots$ & $26.0\pm2.8$ & Daddi+\citeyear{daddi_2010} & 10.99  & 8.75 \\
BzK-16000  & 12      36      30.09   &   62      14      27.96  & 1.522  & 2$-$1  & $0.46\pm0.05$ & 194   & $\dots$ & $18.7\pm2.0$ & Daddi+\citeyear{daddi_2010} &  10.98 & 8.75 \\
BzK-17999  & 12      37      51.82   &   62      15      20.11  & 1.414  & 2$-$1  & $0.57\pm0.06$ & 440   & $\dots$ & $20.1\pm2.1$ & Daddi+\citeyear{daddi_2010} &  11.18 & 8.82 \\
BzK-12591  & 12      37      41.37   &   62      12      51.18  & 1.600  & 2$-$1  & $0.84\pm0.16$ & 600   & $\dots$ & $37.5\pm7.1$ & Daddi+\citeyear{daddi_2010} & 11.32  & 8.79 \\
HDF242    & 12      37      07.20   &   62      14      08.08  & 2.490  & 1$-$0  & $0.21\pm0.03$ & 850   & $\dots$ & $62.5\pm8.9$ & Ivison+\citeyear{ivison_2011} & 11.41  & 8.80 \\
GN20         & 12      37      11.87   &   62      22     12.46  & 4.050  & 2$-$1  & $0.64\pm0.16$ & 670 & $\dots$ & $142.1\pm35.5$ & Carilli+\citeyear{carilli_2010}  &11.40   & 8.80 \\
\hline
\end{tabular}
\begin{list}{}{}
\item[\textbf{Notes.} ]
\item[$^{\mathrm{a}}$]  Significance of our PdBI detections: 1 - secure detection; 2 - tentative detection; 3 - upper limit.  
\item[$^{\mathrm{b}}$] Metallicity are estimated using the mass-metallicity relation and converted into the \citet{denicolo_2002} system.
\end{list}
\end{table*}
\section{Data}
\label{data}
\subsection{\textit{Herschel} and multi-wavelength observations\label{subsec: herschel}}

The new IRAM PdBI CO targets were selected from the \textit{Herschel} observations of the GOODS-N field taken as part of the PACS Evolutionary Probe \citep[PEP\footnote{http://www.mpe.mpg.de/ir/Research/PEP};][]{lutz_2011} guaranteed time key programme.
However, to characterize their FIR/submm SEDs we use here the \textit{Herschel} observations of the GOODS-N field taken as part of the GOODS-\textit{Herschel} key programme \citep[PI: D. Elbaz;][]{elbaz_2011}.
This key programme provides deeper PACS observations of the GOODS-N field along with deep SPIRE observations, allowing us to make use of the full wavelength coverage (100$-$500$\,\mu$m) offered by the \textit{Herschel} Space Observatory.
The GOODS-\textit{Herschel} survey and data reduction methods are described in \citet{elbaz_2011}.
Here, we only summarise the properties relevant for our study.\\
\indent{
PACS and SPIRE source extraction was done using a point-spread-function (PSF) fitting analysis \citep{magnelli_2009}, guided using the position of sources detected in the deep GOODS-N 24 $\,\mu$m observations from the Multiband Imaging Photometer \citep[MIPS;][]{rieke_2004} onboard the \textit{Spitzer} Space Observatory \citep[3$\sigma$$\,\thicksim\,$20$\,\mu$Jy;][]{magnelli_2011a}.
This method both mitigates blending issues encountered in dense fields, and provides a straightforward association between MIPS, PACS and SPIRE sources as well as with the IRAC (Infrared Array Camera) sources from which the MIPS-24/70$\,\mu$m catalogue was constructed \citep{magnelli_2011a}.
The GOODS-N PACS and SPIRE catalogues reach a 3$\sigma$ limit of 1.1, 2.7, 7.5, 9.0 and 12.0 mJy at 100, 160, 250, 350 and 500$\,\mu$m, respectively.
\\}
\indent{
The GOODS-N field benefits from an extensive multiwavelength coverage.
A PSF-matched catalogue was created for GOODS-N as part of the PEP survey\footnote{publicly available at http://www.mpe.mpg.de/ir/Research/PEP/} \citep{berta_2010,berta_2011}, including photometry from far-UV to near-infrared and a collection of spectroscopic redshifts \citep[mainly from][]{barger_2008}, complemented when needed with photometric redshifts computed using EAZY \citep{brammer_2008}.
Stellar masses were derived by fitting \citet{bruzual_2003} templates to these far-UV to near-infrared photometries (see Section \ref{subsec: stellar masses} for more detail).
We cross-matched our IRAC-MIPS-PACS-SPIRE catalogue to this multi-wavelength catalogue, using the IRAC positions of each catalogue and a matching radius of 1\arcsec.
\\}
\indent{
Galaxies with CO observations from the literature were also cross-matched with the UV-to-far-infrared multi-wavelength catalogue. We verified that the available photometry and spectroscopic redshift from the literature were consistent with those from our multi-wavelength catalogue.  We find no systematic discrepancies. 
Finally, we cross-matched using a 9\arcsec matching radius our sample with the 1.16 mm catalogue of \citet{penner_2011}, which was built by combining the 1.1 mm AzTEC \citep{perera_2008} and the 1.2 mm MAMBO \citep{greve_2008} maps.  We found a clear 1.16 mm detection for GN20 and obtained meaningful 1.16 mm upper limits (i.e., 3$\sigma$ flux limits at the location of our galaxies) for the rest of our sample.
Table \ref{tab:herschel} gives the infrared photometry for all the galaxies in the combined sample. 
}
\begin{table*}
\scriptsize
\caption{\label{tab:herschel} Mid- and far-infrared properties of our combined sample}
\centering
\begin{tabular}{ c cc  c c c c c c c c c c} 
\hline \hline
Galaxy & $S_{24}$ & $S_{70}^{\mathrm{a}}$ & $S_{100}^{\mathrm{a}}$ & $S_{160}^{\mathrm{a}}$ & $S_{250}^{\mathrm{a}}$ & $S_{350}^{\mathrm{a}}$ & $S_{500}^{\mathrm{a}}$ & $S_{850}^{\mathrm{b}}$ & $S_{1160}^{\mathrm{c}}$ & $S_{2200}^{\mathrm{d}}$ & $S_{3300}^{\mathrm{d}}$ & $S_{6600}^{\mathrm{d}}$\\
 & {$\mu$Jy} & {mJy} & { mJy} & { mJy} & { mJy} & { mJy} & { mJy} & { mJy} & { mJy} & { mJy} & { mJy} & { mJy}  \\
\hline
PEPJ123712+621753           &$    153\pm5      $ &$ <2.0              $&$  2.6\pm0.3   $&$     3.5   \pm     0.7    $&$    6.2    \pm    2.5    $&$   <9.0  $&$    <12.0 $&$   \dots   $&$  <1.7    $&$ \dots  $&$  \dots    $&$ \dots $ \\
PEPJ123709+621507           &$    128\pm6      $ &$ <2.0              $&$  2.1\pm0.4  $&$      3.0   \pm     1.3   $&$     8.0   \pm     2.5   $&$     <9.0   $&$     <12.0   $&$     \dots   $&$  <1.6    $&$ \dots  $&$  \dots    $&$ \dots $ \\
PEPJ123759+621732           &$    217\pm7      $ &$ <2.4              $&$  1.2\pm0.4    $&$    5.3    \pm    1.7   $&$     10.0  \pm    2.5    $&$    <9.0   $&$   <12.0 $&$   \dots   $&$  <2.1   $&$ \dots  $&$  \dots    $&$ \dots $ \\
PEPJ123721+621346          &$    238\pm8      $ &$ 1.8\pm0.6     $&$  3.1\pm0.3   $&$     8.5    \pm    0.8   $&$    14.5  \pm      2.5   $&$    10.8   \pm     3.0    $&$    8.0   \pm     5.0   $&$    \dots   $&$  <1.6    $&$ \dots  $&$  \dots    $&$ \dots $ \\
PEPJ123615+621008           &$    327\pm7      $ &$ 4.5\pm0.6    $&$  4.3\pm0.4   $&$     9.2  \pm      1.0   $&$    15.9 \pm       3.1   $&$    14.3  \pm      3.0   $&$   <12.0 $&$   \dots   $&$  <1.7    $&$ \dots  $&$  \dots    $&$ \dots $ \\
PEPJ123634+620627           &$    108\pm 9      $ &$ <2.0             $&$  7.1\pm0.4   $&$    13.4  \pm      1.2  $&$     18.1  \pm      4.0   $&$    16.1  \pm      3.0  $&$     15.1    \pm    4.0   $&$    \dots   $&$  <2.0    $&$ \dots  $&$  \dots    $&$ \dots $ \\
PEPJ123633+621005           &$    572\pm7      $ &$ 2.7\pm0.6    $&$  8.5\pm0.3   $&$    16.3 \pm       1.0  $&$     20.2  \pm      2.5   $&$    13.0 \pm       3.0  $&$    <12.0 $&$   \dots   $&$  <1.6    $&$ \dots  $&$  \dots    $&$ \dots $ \\
PEPJ123646+621141           &$    300\pm6      $ &$ <2.0              $&$  2.9\pm0.3   $&$     5.8   \pm     0.7  $&$     12.7   \pm     2.5    $&$   10.7  \pm      3.0  $&$    <12.0   $&$     \dots   $&$  <1.7    $&$ \dots  $&$  \dots    $&$ \dots $ \\
PEPJ123750+621600           &$    188\pm6      $ &$ <2.0              $&$  1.7\pm0.3   $&$     4.8   \pm     1.1    $&$    8.6  \pm      2.5    $&$    8.8  \pm      3.0    $&$    <12.0   $&$     \dots   $&$  <1.8    $&$ \dots  $&$  \dots    $&$ \dots $ \\ 
HDF169      &$    445\pm7      $ &$ 13.2\pm0.7  $&$ 34.2\pm0.6  $&$     63.2   \pm     1.6   $&$    62.5    \pm    3.7  $&$     47.1  \pm      3.0   $&$    19.6   \pm     4.0   $&$     2.2   \pm     0.8  $&$    <1.8    $&$ \dots  $&$  \dots    $&$ \dots $ \\
BzK-4171   &$   140\pm6      $ &$ <3.2               $&$  2.9\pm0.3   $&$     9.0   \pm     1.0   $&$    14.5  \pm     2.5    $&$    <9.0   $&$     <12.0  $&$      \dots   $&$   <1.7   $&$ \dots  $&$  \dots    $&$ \dots $ \\
 BzK-21000 &$   382\pm6      $ &$ 4.9\pm0.7    $&$  8.1\pm0.6   $&$    15.1  \pm      1.4   $&$    24.4  \pm      2.5  $&$     20.1   \pm     4.7   $&$    11.6    \pm    7.4   $&$     \dots   $&$  <2.2    $&$0.32\pm0.15$ &$ 0.04\pm0.06$\\
BzK-16000  &$   190\pm6      $ &$ <2.0              $&$  1.6\pm0.6   $&$     3.7   \pm     0.9  $&$      9.2   \pm     2.5   $&$     <9.0  $&$    <12.0 $&$   \dots   $&$  <1.7    $&$ \dots  $&$  \dots    $&$ \dots $ \\
 BzK-17999 &$   229\pm8      $ &$ <2.0              $&$  4.1\pm0.5   $&$    10.8   \pm     1.4   $&$    14.7  \pm      2.5   $&$    11.2   \pm     3.0    $&$  <12.0 $&$   \dots   $&$  <1.8    $&$ \dots  $&$  \dots    $&$ \dots $ \\
 BzK-12591 &$   377\pm6      $ &$ <3.3              $&$  9.8\pm0.4   $&$    18.3  \pm      1.1  $&$     24.9 \pm       2.5  $&$     20.5  \pm      3.0   $&$     9.9    \pm    4.0   $&$     \dots   $&$  <1.7    $&$ \dots  $&$  \dots    $&$ \dots $ \\
    HDF242   &$   255\pm9      $ &$ <2.0              $&$  1.1\pm0.3    $&$    6.4   \pm     1.3   $&$    23.9   \pm     4.6   $&$    28.3  \pm      6.4  $&$     23.8    \pm    4.0   $&$     8.0   \pm     3.1      $&$    <1.6    $&$ \dots $ &$  \dots    $&$ \dots $ \\
    GN20        &$   68\pm4        $ &$ <2.0              $&$  0.7\pm0.4     $&$   5.4    \pm    1.0   $&$    18.6  \pm      2.7   $&$    41.3  \pm      5.2   $&$    39.7    \pm    6.1   $&$    20.3   \pm     2.0     $&$    10.5    \pm    0.7    $&$ 0.9\pm0.15 $&$ 0.33\pm0.06 $&$ 0.027\pm0.027$ \\
\hline
\end{tabular}
\begin{list}{}{}
\item[\textbf{Notes.} ]
\item $S_{{\rm XX}}$ correspond to flux densities at the observed wavelength XX in microns.
\item [$^{\mathrm{a}}$] Upper limits corresponds to 3$\sigma$ upper limits.
\item [$^{\mathrm{b}}$] These flux densities are taken from \citet{pope_2006}.
\item [$^{\mathrm{c}}$] These flux densities are taken from \citet{penner_2011}.
\item [$^{\mathrm{d}}$] These flux densities are taken from \citet{magdis_2011b}.
\end{list}
\end{table*}
\subsection{IRAM PdBI observations\label{subsec:CO}}
Observations of the 9 {\it Herschel-}selected normal star-forming galaxies were performed at the IRAM PdBI \citep{guilloteau_1992} in 2010 July-October under average summer conditions. 
We observed the CO(2$-$1) line (rest frequency of 230.538 GHz), which is redshifted into the 3mm observing band for sources at $z\sim1$. 
At the time of observing, the PdBI array was in compact configuration with 5 antennae available. 
The WideX correlator was used, providing dual polarization data with spectral resolution of 1.95 MHz over a total bandwidth of 3.6 GHz.

\begin{figure*}
\includegraphics[width=9.cm]{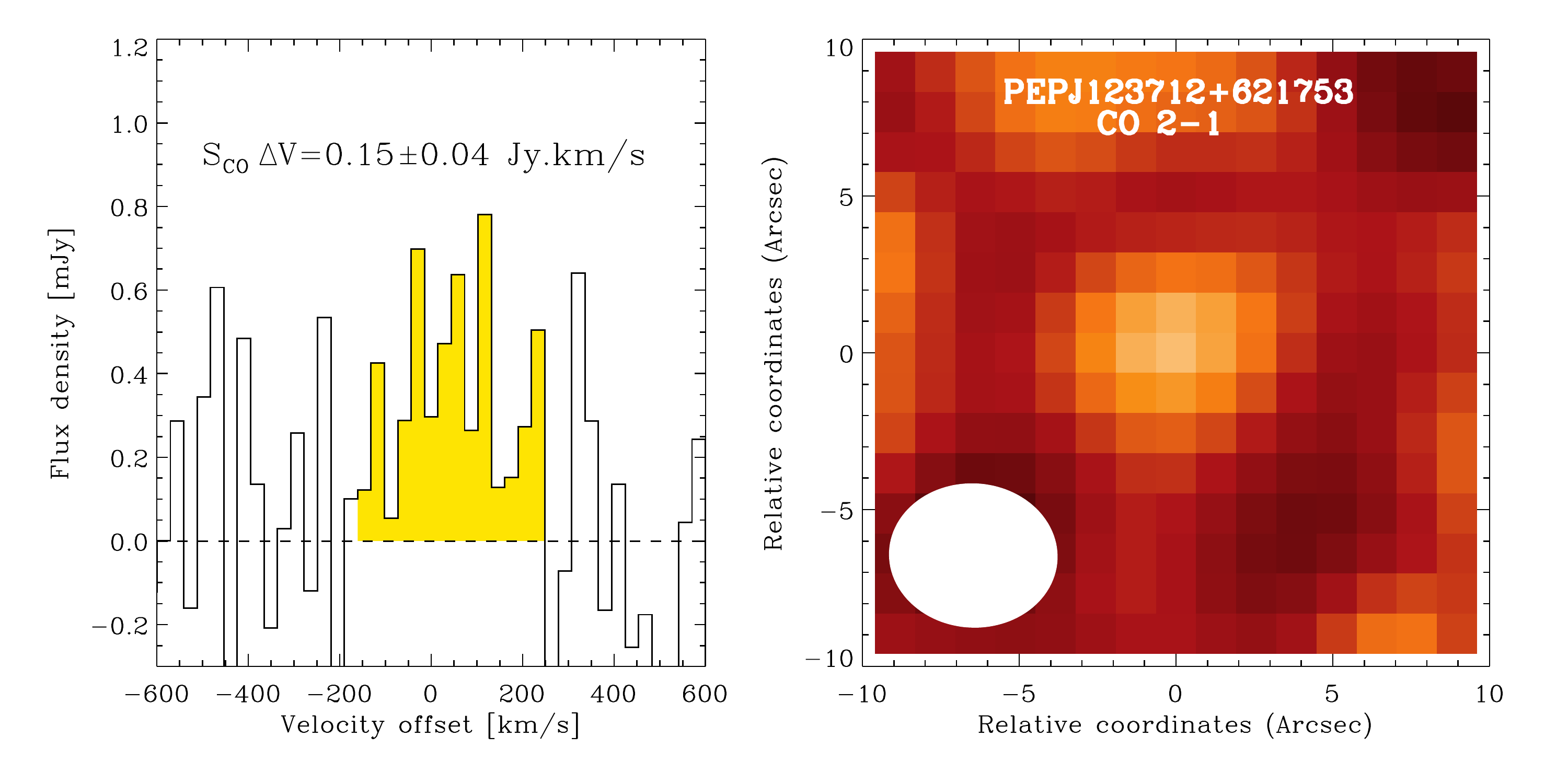}
\includegraphics[width=9.cm]{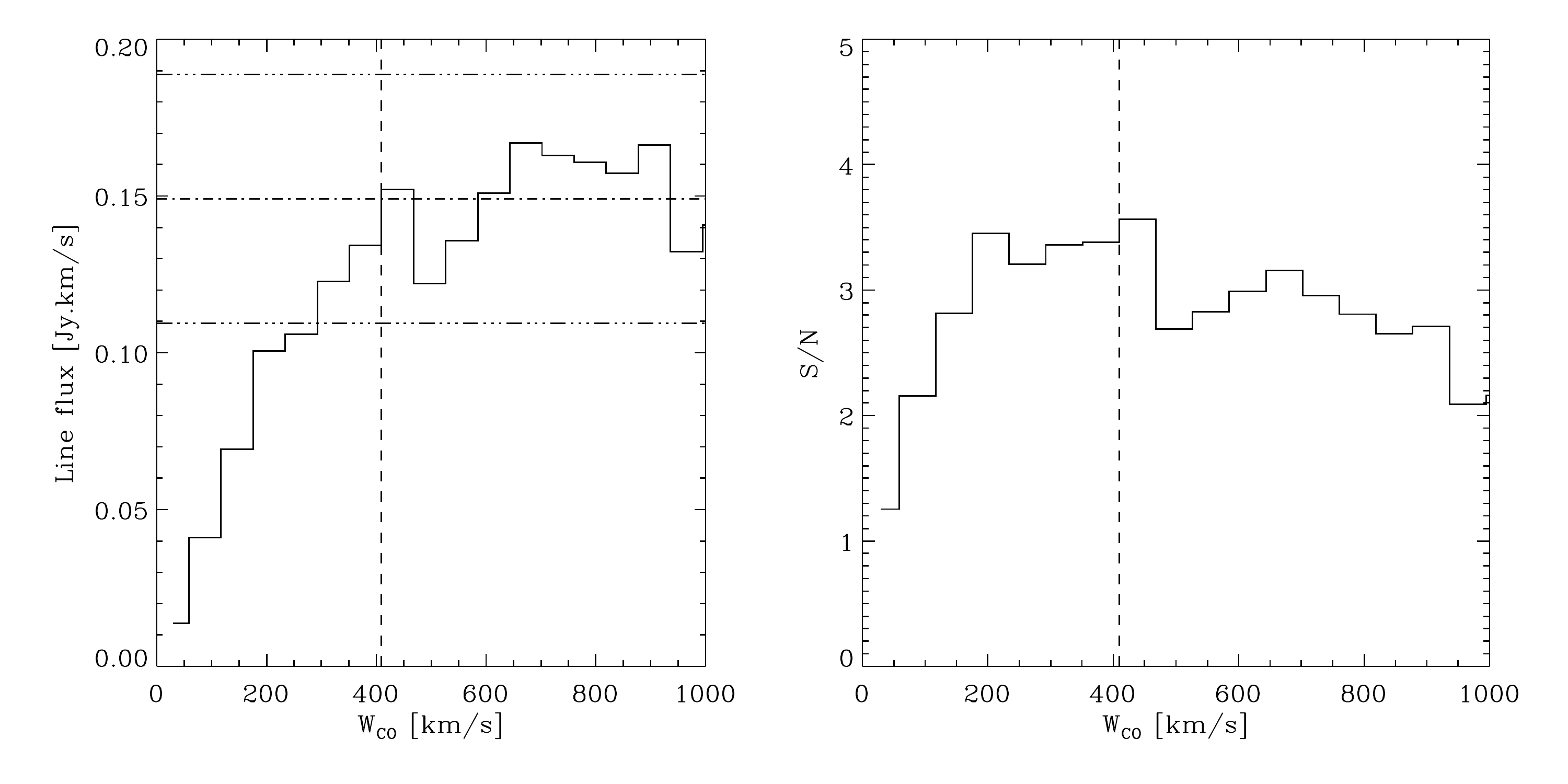}
\includegraphics[width=9.cm]{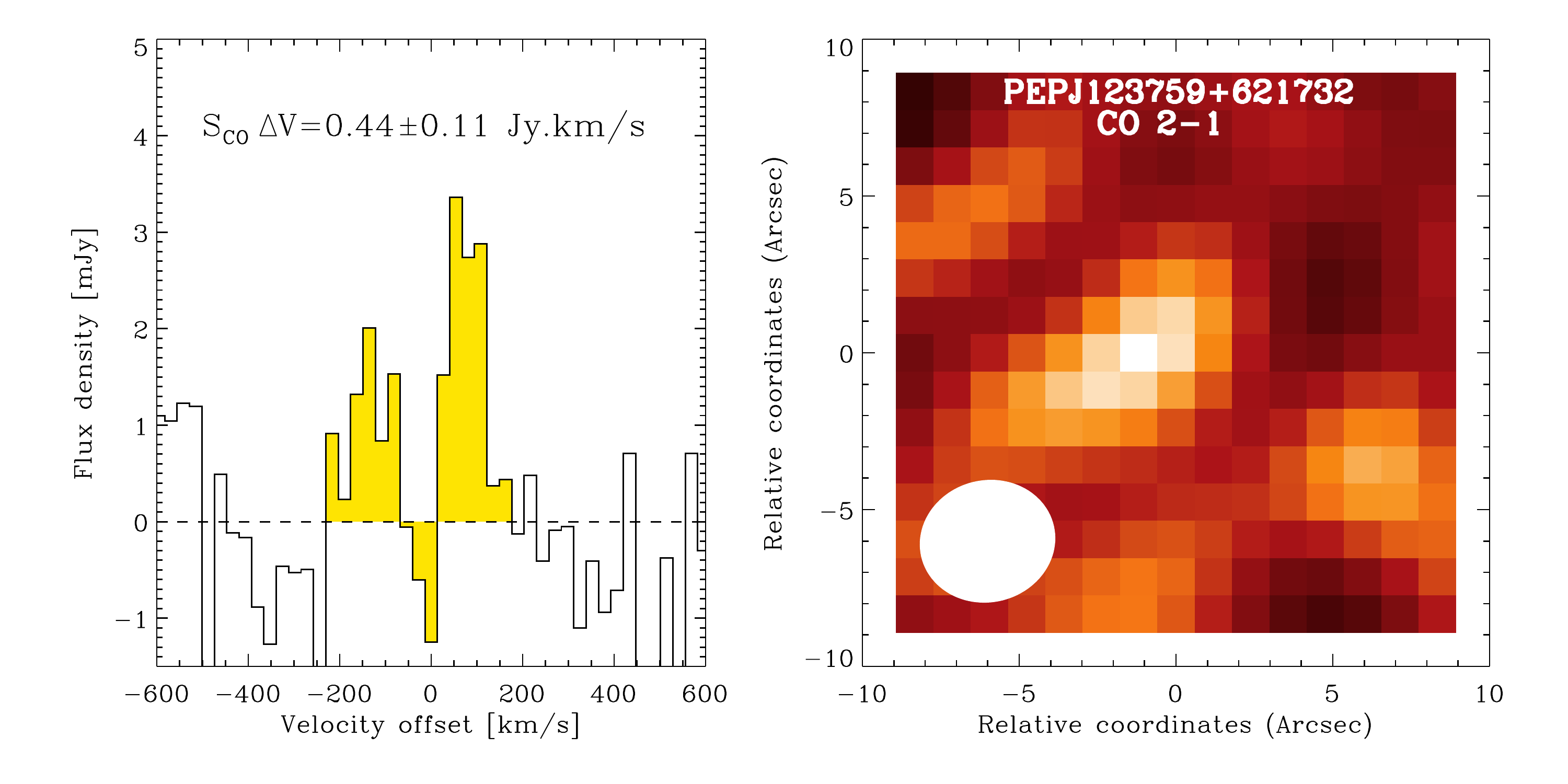}
\includegraphics[width=9.cm]{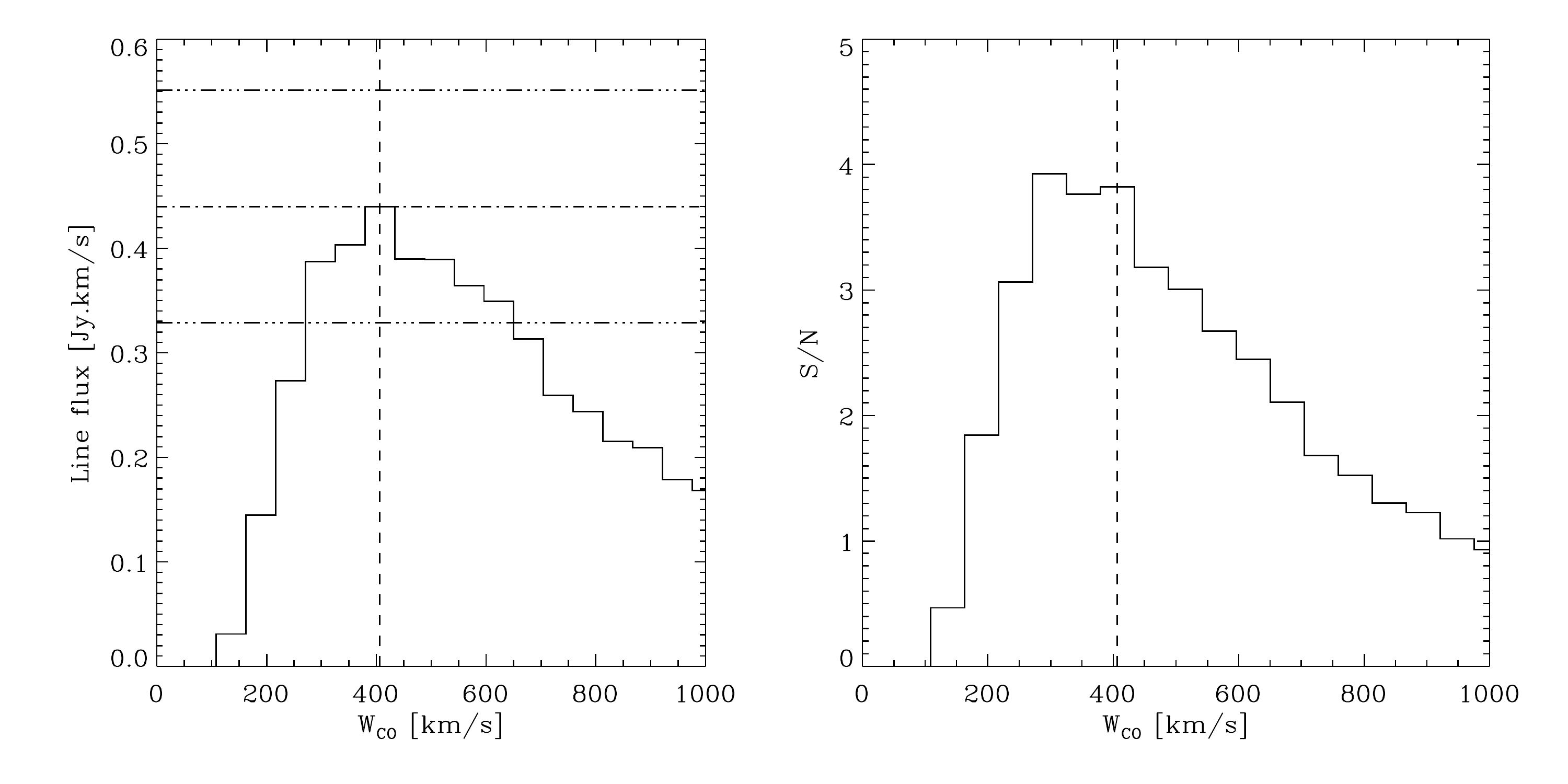}
\includegraphics[width=9.cm]{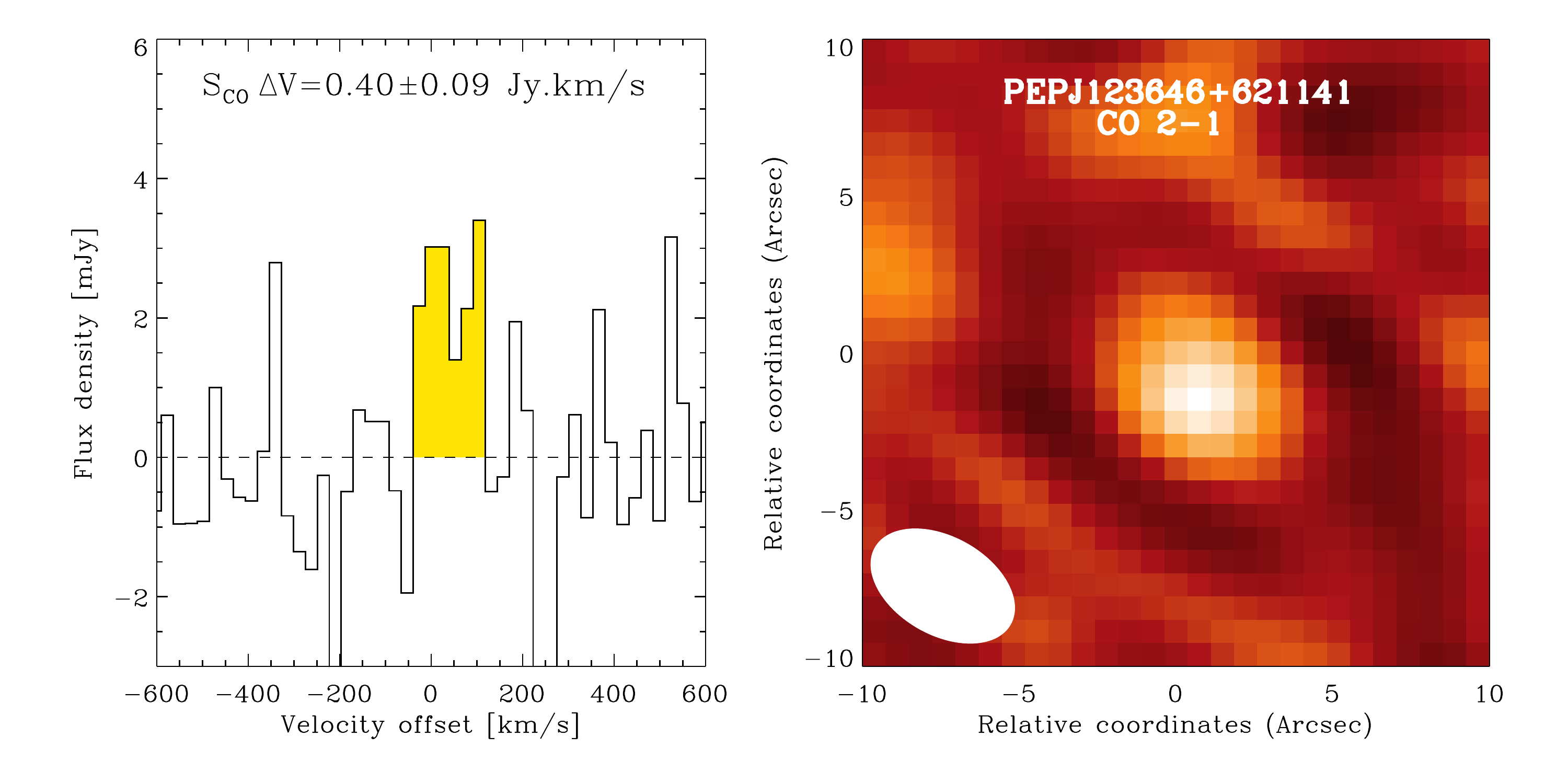}
\includegraphics[width=9.cm]{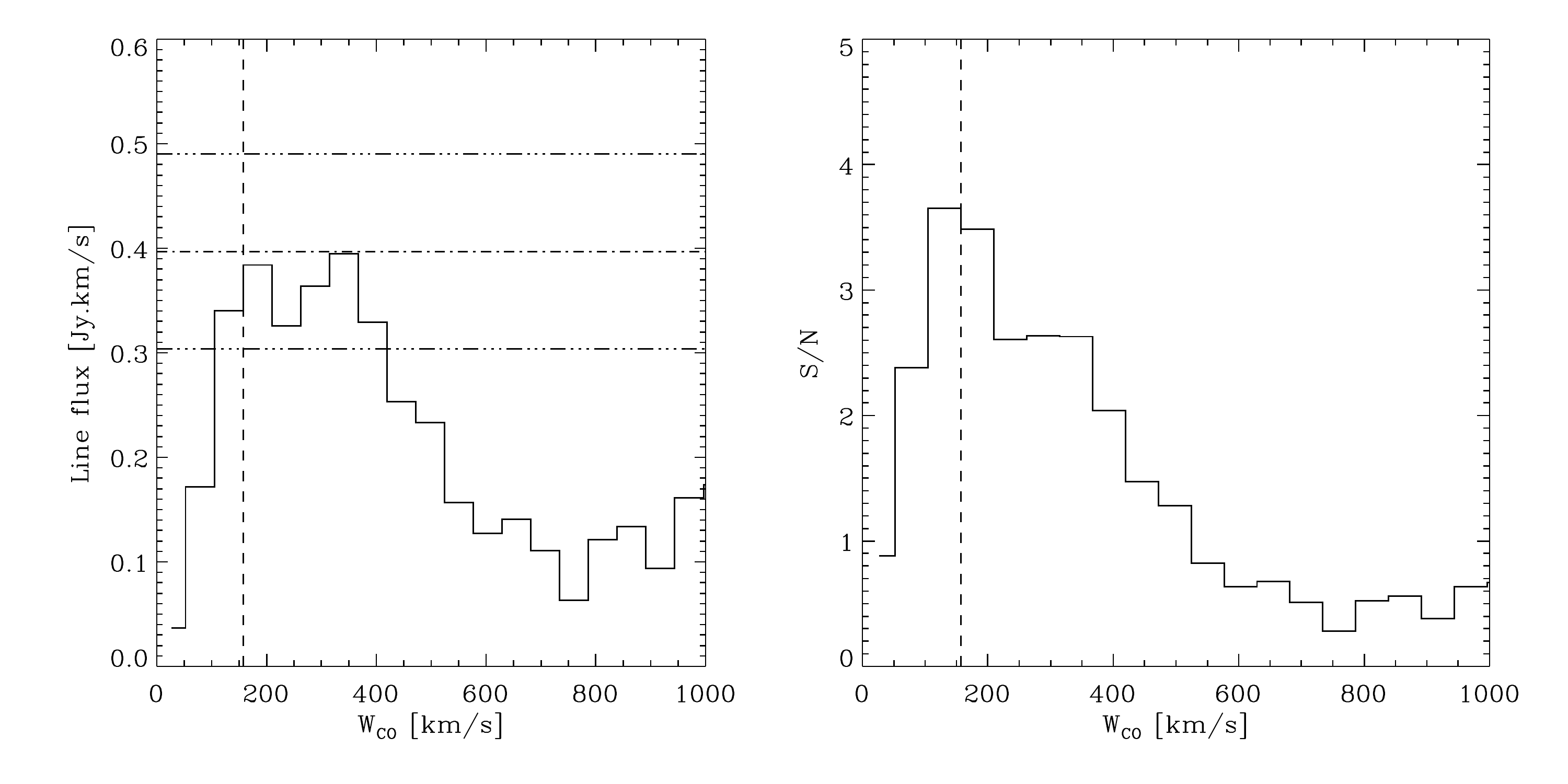}
\includegraphics[width=9.cm]{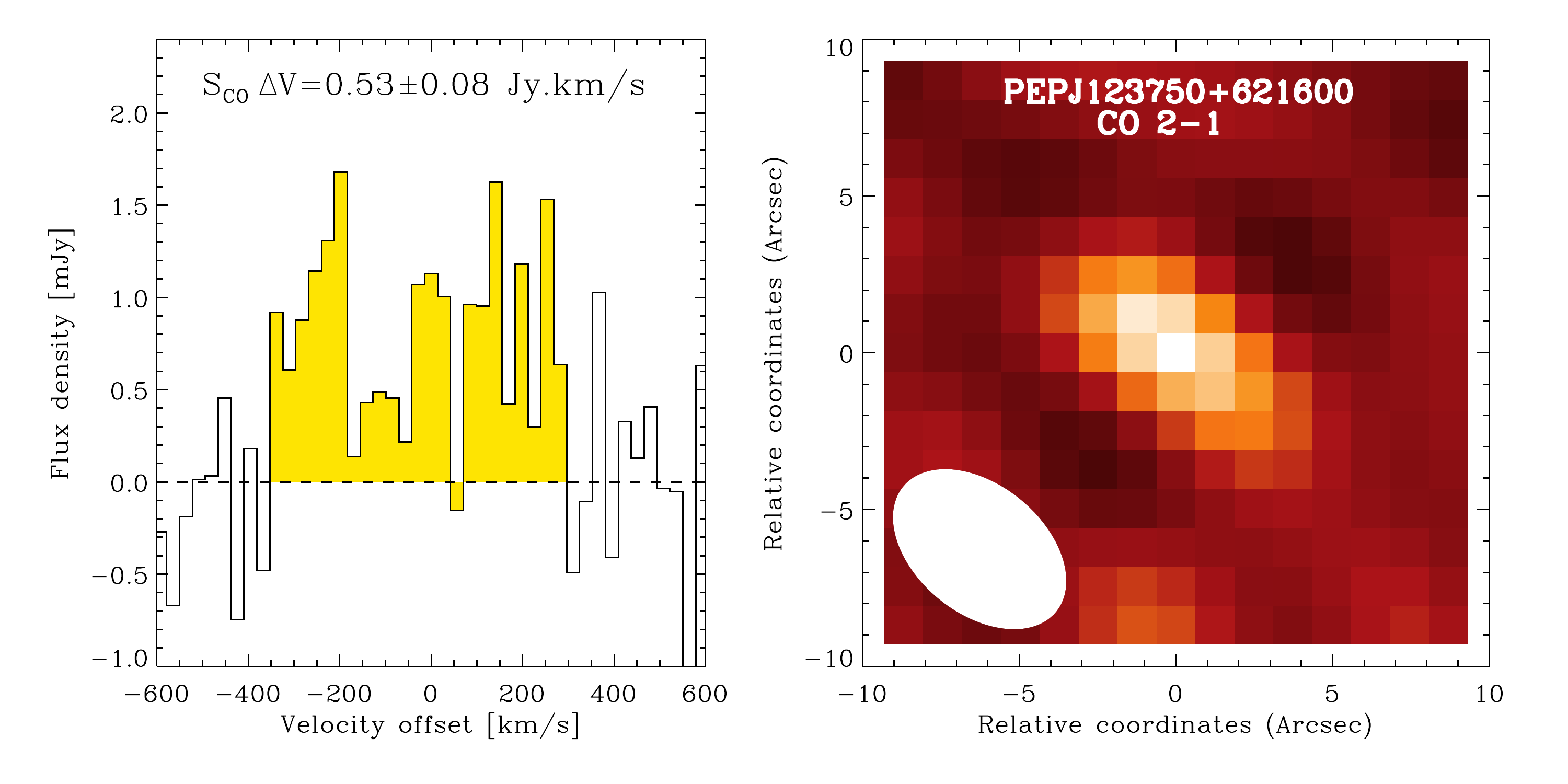}
\includegraphics[width=9.cm]{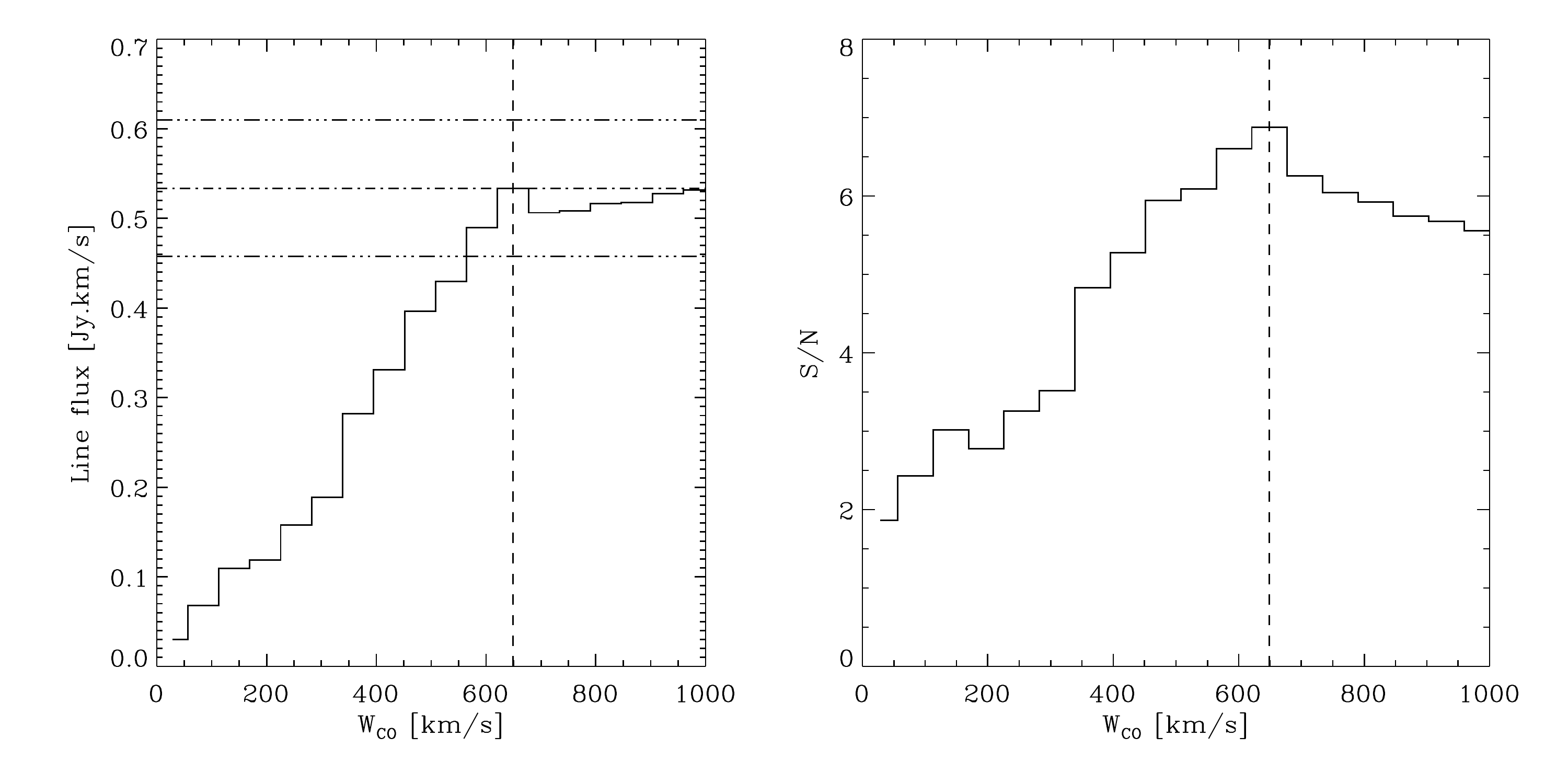}
\includegraphics[width=9.cm]{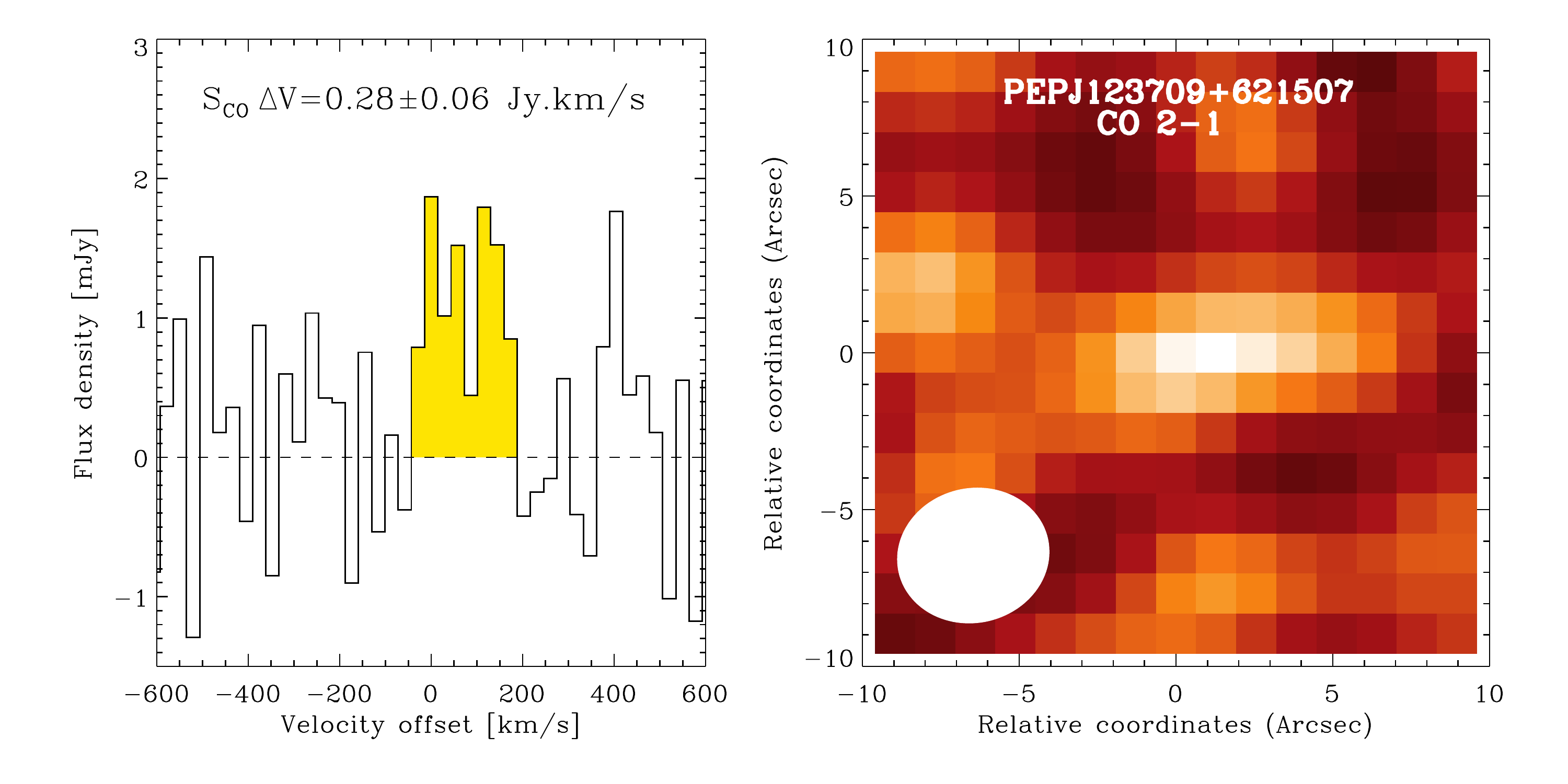}
\hspace{0.3cm}\includegraphics[width=9.cm]{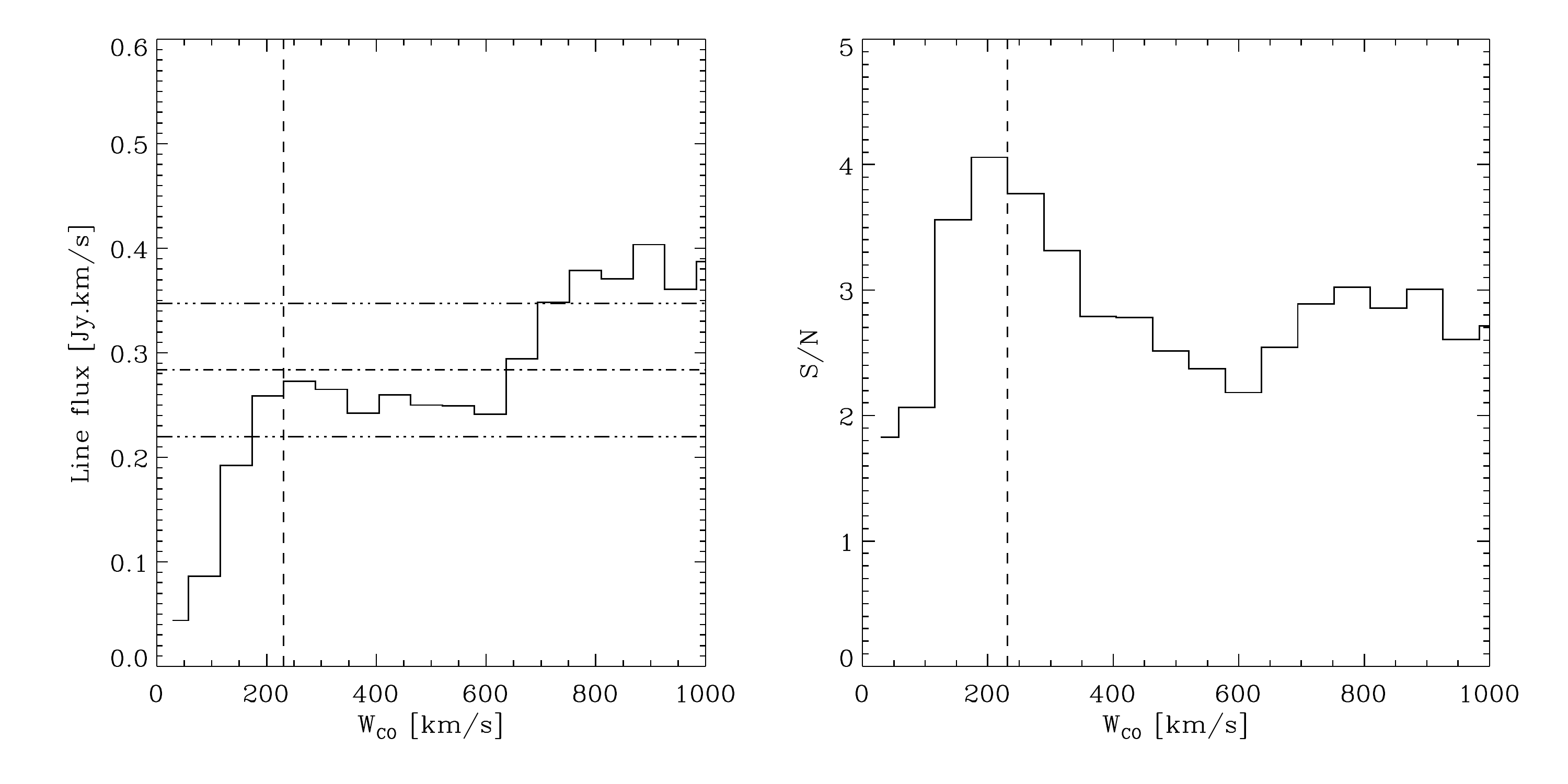}
\caption{CO(2-1) spectrum (\textit{left}), integrated line map (\textit{center left} panel), line flux versus $W_{{\rm CO}}$ (\textit{center right} panel) and signal over noise (S/N) versus $W_{{\rm CO}}$ (\textit{right} ) of our 7 PdBI targets with secure (PEPJ123712+621753, PEPJ123759+621732, PEPJ123646+621141 and PEPJ123750+621600) and tentative line detection (PEPJ123709+621507, PEPJ123721+621346 and PEPJ123633+621005).
In each spectrum, the yellow area displays the part of the spectrum over which the line fluxes are integrated.
Dashed lines in the right two panels outline this value of $W_{{\rm CO}}$.
To create the center right and right panels, we fixed the central frequency when increasing $W_{{\rm CO}}$ to the central frequency of the yellow area displayed in the left panel.
Dotted-dashed lines and triple-dotted-dashed lines represent the line flux and error of each object, respectively.
Beam sizes are displayed in the lower left corner of each integrated line map.
\label{Spectra}}
\end{figure*}

\begin{figure*}
\includegraphics[width=9.cm]{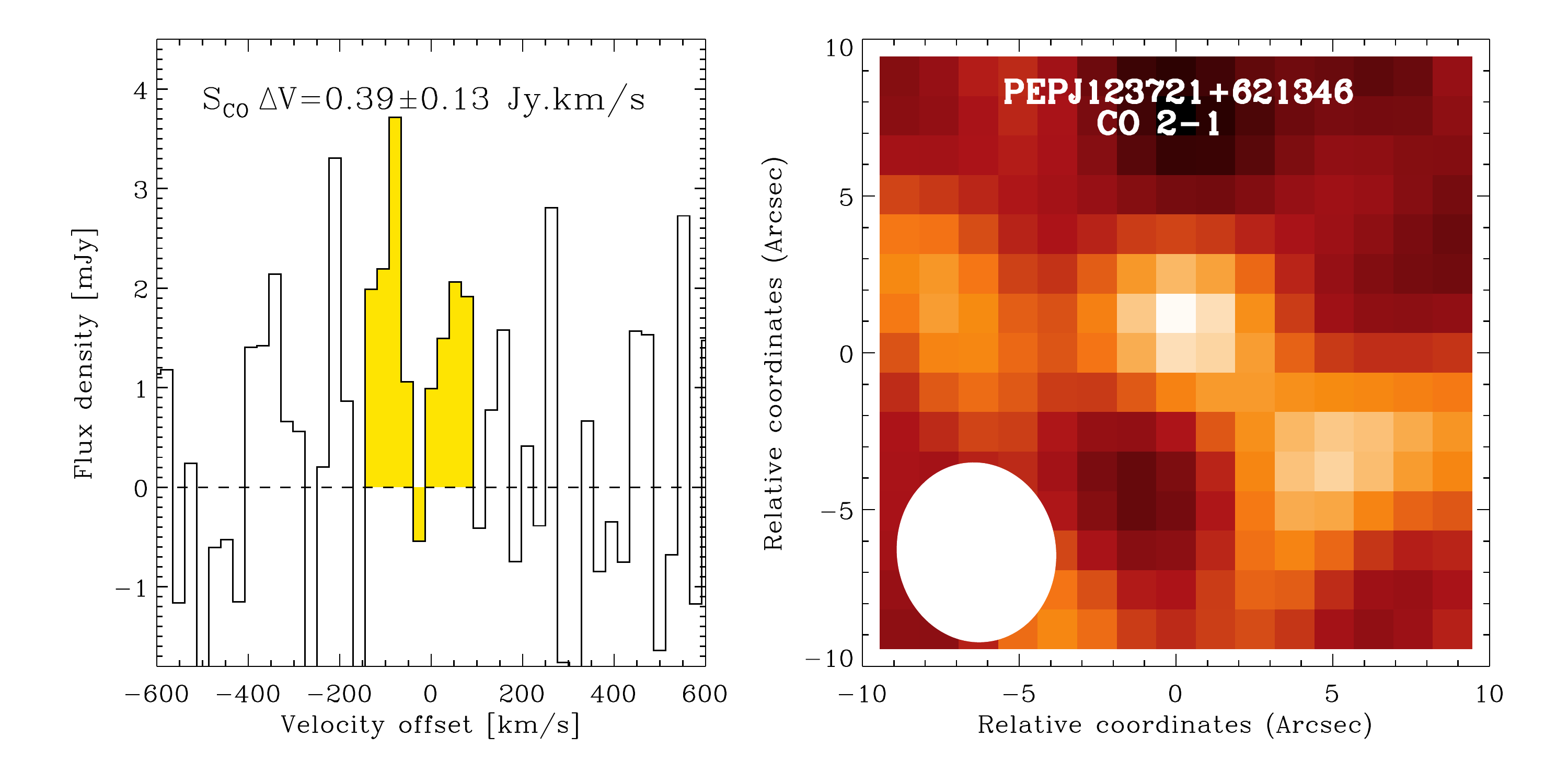}
\includegraphics[width=9.cm]{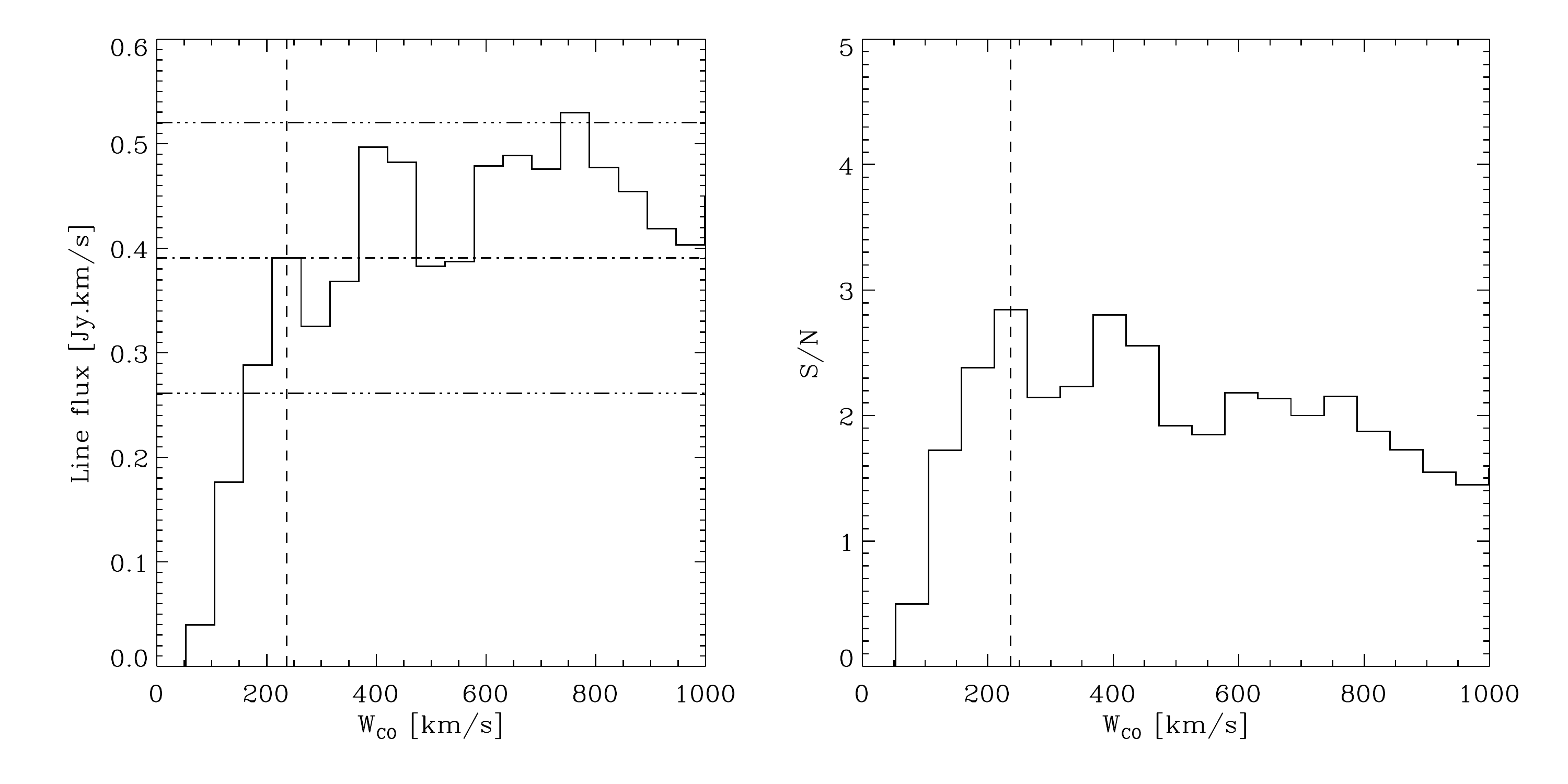}
\includegraphics[width=9.cm]{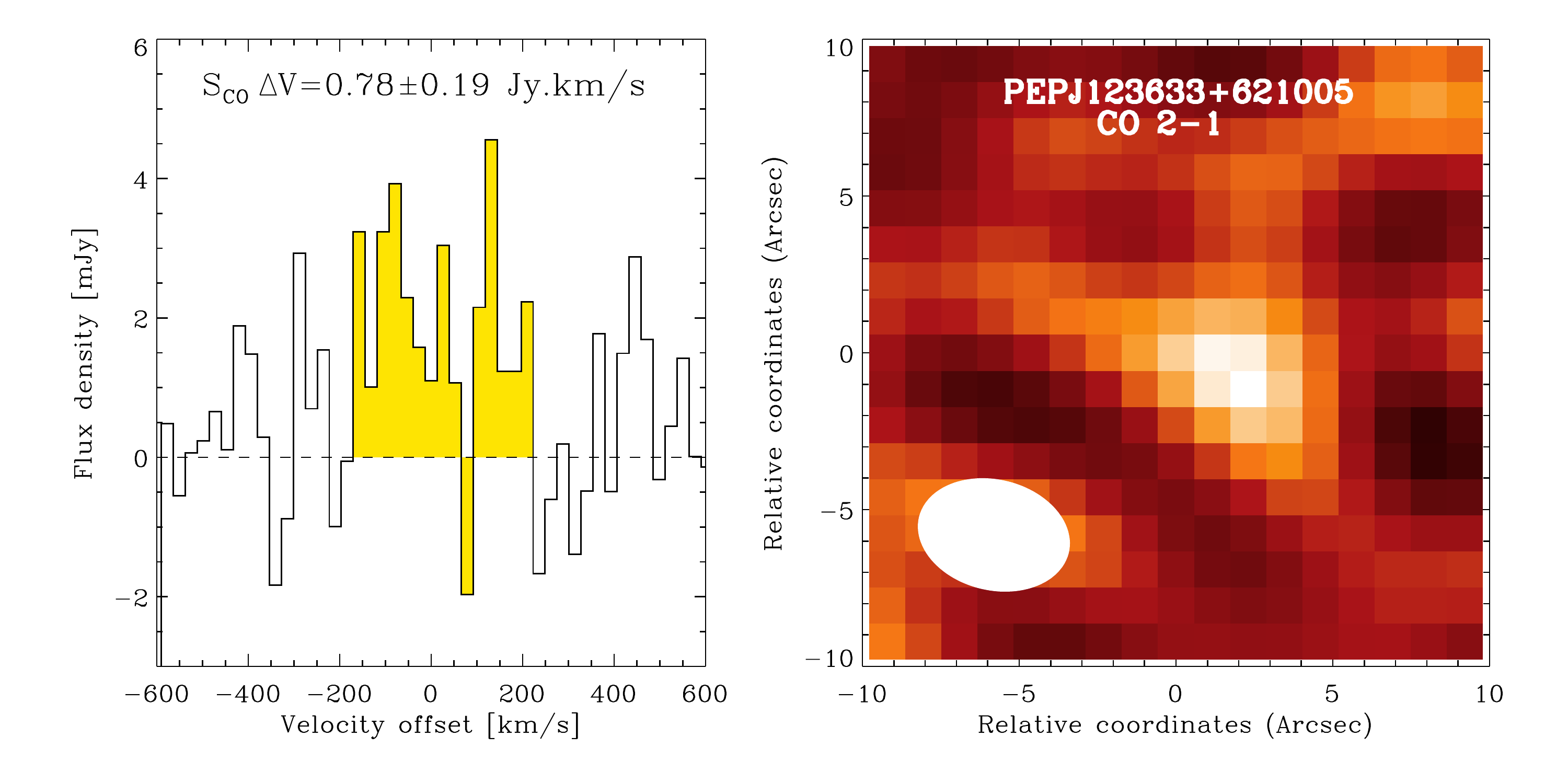}
\hspace{0.3cm}\includegraphics[width=9.cm]{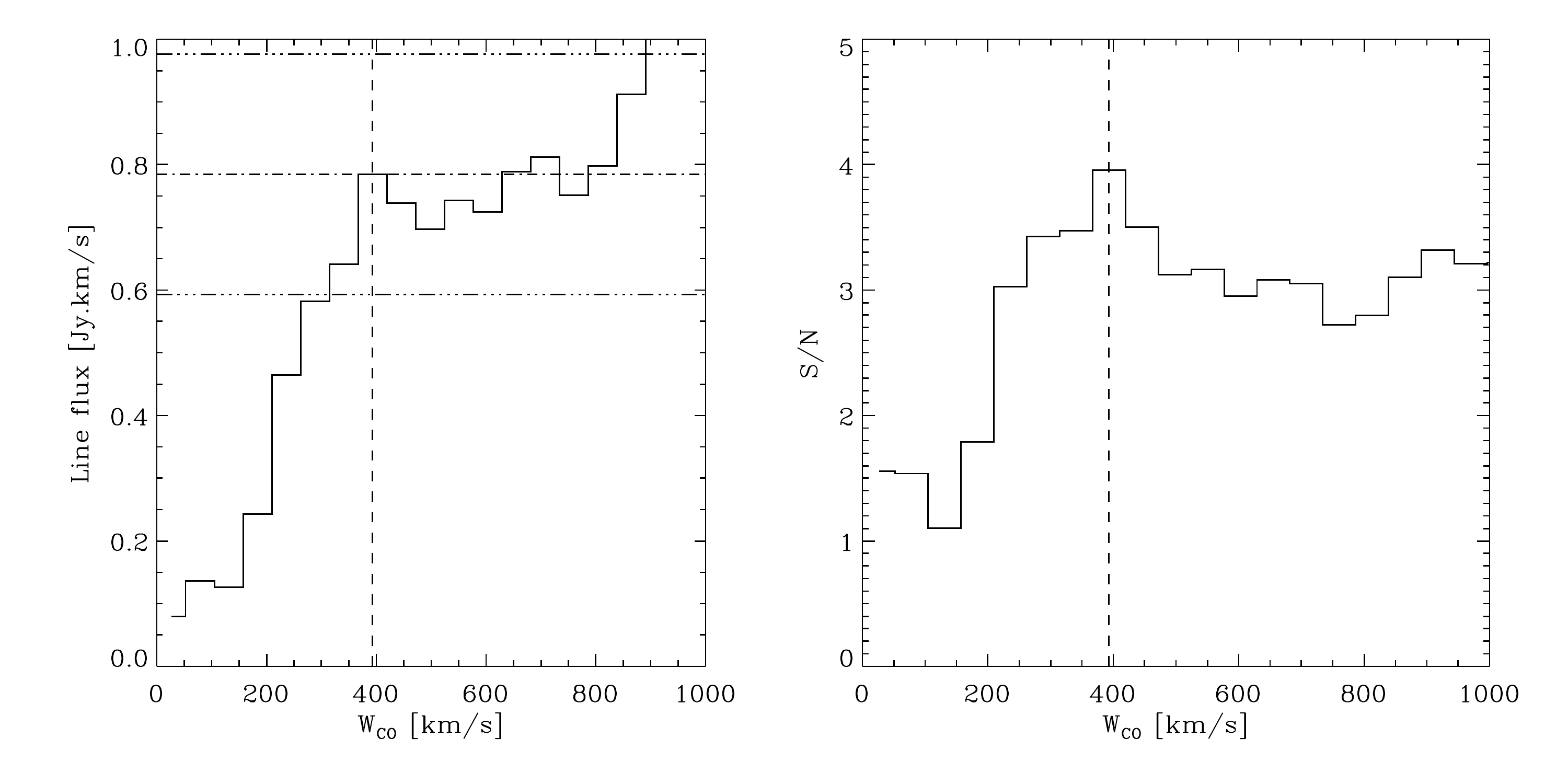}
\caption{Figure \ref{Spectra} continued
\label{Spectra suite}}
\end{figure*}

The data were calibrated with CLIC within the standard IRAM software package, GILDAS\footnote{http://www.iram.fr/IRAMFR/GILDAS}.
A standard passband calibration scheme was applied, followed by phase and amplitude calibration.
The flux calibration was done using observation of reference bright quasars.
Due to the poor observing conditions during some of the runs (e.g. high precipitable water vapour, strong winds, low elevation), particular care was taken to flag data with high phase noise.
Finally, data cubes were created using the MAPPING environment and cleaned using the CLARK version of CLEAN implemented in GILDAS.
The typical RMS in the data cubes is of 0.8 mJy/beam with a typical beam diameter of $\thicksim\,$5\arcsec.
The data cubes were examined for sources at the expected spatial and spectral positions.
The integrated line map and spectrum of each target are shown in Figures \ref{Spectra} and \ref{Spectra suite}.
Four of the sources are securely detected, three are tentative detections, and the remaining two are non-detections.
For the detections, line fluxes were measured by integrating the spectra over the width of the signal ($W_{{\rm CO}}$).
Due to the absence of continuum in our spectra, $W_{{\rm CO}}$ is defined as the frequency range maximizing the signal to noise ratio (S/N) (i.e., the peak of the S/N versus $W_{{\rm CO}}$ plot). This $W_{{\rm CO}}$ also corresponds to the frequency range where the curve of growth of the line flux reaches its plateau (see center right and right panels of Figure \ref{Spectra} and \ref{Spectra suite}).
To create the S/N versus $W_{{\rm CO}}$ and line flux versus $W_{{\rm CO}}$ diagrams, we have to fix the central frequency around which we increase $W_{{\rm CO}}$.
These fixed frequencies are defined as being the central frequencies of the yellow areas displayed in the left panels of Fig. \ref{Spectra} and \ref{Spectra suite}.
Uncertainties on $W_{{\rm CO}}$ are defined as being the range of $W_{{\rm CO}}$ over which S/N is greater than max[S/N($W_{{\rm CO}}$)]$-1$.
For non-detections we measured 3$\sigma$ upper limits assuming $W_{{\rm CO}}$$\,=\,$300$\,$km$\,$s$^{-1}$, typical for high-redshift galaxies \citep{tacconi_2010}. A summary of the observed CO fluxes and $W_{{\rm CO}}$ for both literature and new sources is presented in Table \ref{tab:pdbi}. 


\section{Analysis}
\label{analysis}
\subsection{The CO(1-0) luminosities}
Since the standard techniques to extrapolate \mh\ from CO measurements are calibrated for observations of the $J=1 \rightarrow 0$ transition, the fluxes coming from observations done in higher transitions need to be converted into fluxes of the $J=1 \rightarrow 0$ transition.
All but one of the galaxies have observations of the $J=2 \rightarrow 1$ transition, with HDF242 (aka GN19) observed directly in $J=1 \rightarrow 0$ transition \citep{ivison_2011}. 
Thanks to this homogeneity, our combined sample is mostly free of uncertainties in the excitation corrections when combining multi-transition samples.
We adopt a ratio of $S_{(2-1)}/S_{(1-0)}$$\,=\,$$3$, which is typically observed within the disks of local normal star-forming galaxies \citep[e.g.,][]{leroy_2009} as well as in normal high-redshift star-forming galaxies \citep[i.e., \textit{BzK} galaxies; see e.g.,][]{aravena_2010}\footnote{
It corresponds to a $L^{\prime}_{{\rm CO(2-1)}}/L^{\prime}_{{\rm CO(1-0)}}$$\,\equiv\,$$T_{b}(2$$\,\rightarrow\,$$1)/T_{b}(1 \rightarrow 0)$ ratio of 0.75, where $T_{b}$ is the equivalent Rayleigh-Jeans brightness temperature.}.
Recently \citet{bothwell_2012} found $S_{(2-1)}/S_{(1-0)}$$\,=\,$$3.36\pm0.52$ for a sample of 40 luminous SMGs, i.e. galaxies mostly situated far above the MS of star-formation.
Because there is a clear agreement between the $S_{(2-1)}/S_{(1-0)}$ flux ratio found in normal SFGs and SMGs, we conclude that the conversion of CO $J=2 \rightarrow 1$ fluxes into CO $J=1 \rightarrow 0$ fluxes does not introduce any differential biases in the CO(1-0) luminosity estimates of on- and off-MS galaxies.
The CO(1-0) luminosities can then be calculated following \citet{solomon_1997}:
\begin{equation}
L^{\prime}_{{\rm CO(1-0)}}=3.25 \times 10^7 S_{(1-0)} \Delta V\  \nu^{-2}_{(1-0)} D_{{\rm L}}^2 (1+z)^{-1},
\end{equation}
where $S_{(1-0)} \Delta V$ is the integrated line flux in Jy\,km\,s$^{-1}$, $\nu_{(1-0)} \equiv 115.271$ GHz is the rest-frame frequency of the CO(1-0) line, and $D_{{\rm L}}$ is the luminosity distance in Mpc.   
Table \ref{tab:pdbi} summarises the inferred CO properties of our sample.\\
\subsection{Dust masses, dust temperatures and SFRs \label{subsec: dust masses}}
To infer the dust mass of our galaxies we use the dust model of \citet[][hereafter DL07]{draine_2007}.
This model describes the interstellar dust as a mixture of carbonaceous and amorphous silicate grains with a size distribution mimicking the Milky Way (MW) extinction curve.
The smallest carbonaceous grains have the properties of PAHs and their abundance is parametrized by the PAH index, $q_{{\rm PAH}}$, i.e., the fraction of dust mass in the form of PAH grains with less than $10^3$ carbon atoms.
The DL07 model assumes that a large fraction of the dust (i.e., $1-\gamma$ with $0.0<\gamma<0.3$) is situated in the diffuse ISM and exposed to a radiation field with a constant intensity $U_{{\rm min}}$ (U being a dimensionless scale factor normalized to the local interstellar radiation field of the MW).
The remaining fraction of the dust (i.e., $\gamma$) is exposed to a power-law ($\alpha$) radiation field ranging from $U_{{\rm min}}$ to $U_{{\rm max}}$ and corresponds to the dust enclosed in photodissociation regions (PDRs).
Combining these two components, the dust emission spectrum of a galaxy is given by :
\begin{equation}
L_{\nu} = M_{{\rm dust}}\ \big\lbrack (1-\gamma ) p_{\nu}^{(0)} (q_{{\rm PAH}},U_{{\rm min}})\ +\ \gamma p_{\nu}(q_{{\rm PAH}},U_{{\rm min}},U_{{\rm max}},\alpha)\big\rbrack
\end{equation}
where $p_{\nu}^{(0)} (q_{{\rm PAH}},U_{{\rm min}})$ is the emitted spectrum per unit dust mass for the component heated by a single radiation field $U_{{\rm min}}$, and $p_{\nu}(q_{{\rm PAH}},U_{{\rm min}},U_{{\rm max}},\alpha)$ is the emission spectrum for the PDR dust.
DL07 found that the exact value of $\alpha$ does not have a large influence on the quality of their fits and fix $\alpha=2$.
In addition, they suggest to fix $U_{{\rm max}}=10^6$ and to use $0.7\lesssim U_{{\rm min}} \lesssim 25$ when submillimeter data are not available.
This lower cutoff for $U_{{\rm min}}$ is made to avoid erroneously fitting very cold dust components to a set of data with no submm constraints.
DL07 conclude that this cutoff does not introduce any systematic biases in their dust mass estimates.

Following these prescriptions, we build a grid of models with different PAH abundances ($0.47\%<q_{{\rm PAH}}<4.6\%$), values of $U_{{\rm min}}$ ($0.7-25$) and values of $\gamma$ ($0.0-0.3$).  At each grid point, we compare the model SED and the FIR/submm photometry, with the dust mass given by the normalization of the SED that minimizes $\chi^2$.   Integrating this best-fitting SED template from rest-frame 8 to 1000$\,\mu$m, we also infer for the grid point the infrared luminosity ($L_{{\rm IR}}$) and SFR (with SFR[M$_{\odot}$yr$^{-1}$]$\,=\,$$10^{-10}\,L_{{\rm IR}}$[L$_{\odot}$], assuming a Chabrier IMF).  We assign for each galaxy a value of $M_{{\rm dust}}$, $L_{{\rm IR}}$ and SFR given by the geometric mean of the respective quantity across the region of the parameter space where $\chi^2<\chi_{{\rm min}}^2+1$.  These quantities, along with their 1$\sigma$ uncertainties derived over the same grid points, are given in Table \ref{tab: mdust}.

We find that our grid of DL07 models provides a good fit to the photometric data points, with a median $\chi^2=1.8$ for $N_{{\rm dof}}=1$.
This is in agreement with DL07, \citet{munoz-mateos_2009} and \citet{dale_2012}, who found that this particular set of models could reproduce well the large wealth of data available for the SINGS and KINGFISH galaxies.
We note that some of our galaxies are also included in \citep[][; i.e., the BzK galaxies]{magdis_2012}.
Our dust mass estimates for these galaxies are in agreement with those of \citet{magdis_2012} who also used the DL07 model.

Among our 17 galaxies, 12 have clear (S/N$>3$, 8 galaxies) or tentative detections (3$\,>\,$S/N$\,>\,$2, 4 galaxies) of {\it Herschel} fluxes up to the rest-frame 160$\,\mu$m wavelength.
For these galaxies our dust mass estimates should be robust.
Indeed, \citet{dale_2012} found that the lack of $\lambda_{{\rm rest}}>160\,\mu$m observations does not introduce any systematic biases, although the scatter in the resulting dust masses may increase by up to $\thicksim50\%$.  The main conclusions of this paper are based on the comparison of our dust and gas measurements with empirical relations presented in \citet{leroy_2011}, where results are also based on FIR measurements short-ward of rest-frame 160$\,\mu$m.  This consistency with the methodology of \citet{leroy_2011} trumps possible worries regarding the overall normalization of our dust masses in the absence of submm data.

For the remaining 5 galaxies, reliable {\it Herschel} fluxes are only available up to rest-frame $100\,\mu$m.  To test for a systematic bias in the dust mass estimates of these galaxies, we re-fitted the other 12 galaxies this time omitting any flux measurements at $\lambda_{\rm rest}>100\,\mu$m.   We find no systematic biases, although the scatter in the resulting dust masses increases by $\thicksim50\%$ compared to the reference values obtained using the flux measurements at $\lambda_{\rm rest}>100\,\mu$m.

\begin{table}
\scriptsize
\caption{\label{tab: mdust} Dust properties of our combined sample}
\centering
\begin{tabular}{ c c c c c} 
\hline \hline
 \rule{0pt}{2ex}Galaxy & log($M_{{\rm dust}}^{{\rm DL07}}$) & log($L_{{\rm IR}}^{{DL07}}$)  & log($M_{{\rm dust}}^{{\rm BB}}$) & $T_{{\rm dust}}^{{\rm BB}}$\\
              &  {M$_{\odot}$} &  { L$_{\odot}$}  & { M$_{\odot}$} & { K} \\
\hline
PEPJ123712+621753       & $ 8.91 \pm 0.42$ & $11.64 \pm 0.05$ & $ 7.93 \pm 0.57$ & $ 31.5 \pm  6.5 $ \\
PEPJ123709+621507       & $ 8.96 \pm 0.40$ & $11.58 \pm 0.06$ & $ 8.50 \pm 0.60$ & $ 25.0 \pm  5.0 $ \\
PEPJ123759+621732       & $ 8.94 \pm 0.27$ & $11.46 \pm 0.05$ & $ 8.27 \pm 0.28$ & $ 27.0 \pm  3.0 $ \\
PEPJ123721+621346       & $ 9.07 \pm 0.11$ & $11.65 \pm 0.01$ & $ 8.45 \pm 0.10$ & $ 26.0 \pm  1.0 $ \\
PEPJ123615+621008       & $ 9.33 \pm 0.12$ & $11.78 \pm 0.01$ & $ 8.50 \pm 0.10$ & $ 26.0 \pm  1.0 $ \\
PEPJ123634+620627       & $ 8.56 \pm 0.16$ & $11.95 \pm 0.03$ & $ 8.55 \pm 0.10$ & $ 29.0 \pm  1.0 $ \\
PEPJ123633+621005       & $ 8.61 \pm 0.05$ & $11.87 \pm 0.01$ & $ 8.32 \pm 0.07$ & $ 30.0 \pm  1.0 $ \\
PEPJ123646+621141       & $ 9.26 \pm 0.15$ & $11.61 \pm 0.01$ & $ 8.52 \pm 0.12$ & $ 24.0 \pm  1.0 $ \\
PEPJ123750+621600       & $ 8.97 \pm 0.25$ & $11.58 \pm 0.03$ & $ 8.35 \pm 0.25$ & $ 26.5 \pm  2.5 $ \\
HDF169     & $ 9.03 \pm 0.05$ & $12.61 \pm0.01$ & $ 8.60 \pm 0.05$ & $ 37.0 \pm  0.5 $ \\
BzK-4171   & $ 8.65 \pm 0.12$ & $11.97 \pm 0.03$ & $ 8.18 \pm 0.18$ & $ 34.5 \pm  2.5 $ \\
 BzK-21000 & $ 9.30 \pm 0.02$ & $12.38 \pm 0.01$ & $ 8.52 \pm 0.03$ & $ 33.5 \pm  0.5 $ \\
BzK-16000  & $ 8.96 \pm 0.32$ & $11.84 \pm 0.04$ & $ 8.43 \pm 0.43$ & $ 29.0 \pm  4.0 $ \\
BzK-17999 & $ 8.67 \pm 0.11$ & $12.03 \pm 0.02$ & $ 8.25 \pm 0.15$ & $ 34.0 \pm  2.0 $ \\
 BzK-12591 & $ 8.83 \pm 0.05$ & $12.42 \pm0.01$ & $ 8.42 \pm 0.03$ & $ 36.5 \pm  0.5 $ \\
   HDF242 & $ 9.30 \pm 0.06$ & $12.65 \pm 0.01$ & $ 9.02 \pm 0.12$ & $ 33.0 \pm  2.0 $ \\
    GN20   & $ 9.66 \pm 0.01$ & $13.23 \pm 0.01$ & $ 9.15 \pm 0.05$ & $ 41.0 \pm  0.5 $ \\
\hline
\end{tabular}
\end{table}

In order to obtain a proxy of the luminosity-weighted dust temperature of each galaxy, we fit their far-infrared SEDs, in the optically thin approximation, with a single modified blackbody function:
\begin{equation}
S_{\nu}\propto\frac{\nu^{3+\beta}}{{\rm exp}(h\nu/kT_{{\rm dust}})-1},
\end{equation}
where $S_{\nu}$ is the flux density, $\beta$ is the dust emissivity spectral index and $T_{{\rm dust}}$ is the dust temperature.
We fix the dust emissivity spectral index to 1.5, a standard value shown to provide a good fit to the far-infrared SEDs of high-redshift galaxies \citep[see, e.g.,][]{magnelli_2012}.
Since a single temperature model cannot fully describe the Wien side of the far-infrared SED of galaxies, because short wavelength observations are dominated by a warmer or transiently heated dust component, we exclude from our fitting procedure flux measurements with $\lambda_{{\rm rest}}<50\,\mu$m.

The absolute dust temperature derived using a single blackbody model strongly depends on the assumed dust emissivity.
For example, using $\beta=1.5$ leads to dust temperatures systematically higher than if using $\beta=2.0$ ($\Delta T_{{\rm dust}}\,$$\thicksim\,$$4\,$K).
Nonetheless, fixing $\beta$ to 1.5 or allowing it to vary freely between $1.5-2.0$ does not significantly affect the main results presented in Section \ref{alphaco}.   In any case, to prevent our results from only relying on the dust temperature, we also use the far-infrared color of galaxies (i.e., $f_{\nu}[60\,\mu{\rm m}]/f_{\nu}[100\,\mu{\rm m}]$; hereafter $f_{60}/f_{100}$) as an independent measure of the conditions prevailing in their ISMs \citep[see, e.g.,][]{dale_2001}.  
Here, $f_{60}/f_{100}$ is simply obtained using a log-linear interpolation between all our observed far-infrared data points.
Because our observations always encompass the rest-frame 60 and 100$\,\mu$m passbands, these estimates are robust and model independent.

\begin{figure}
\center
\includegraphics[width=9.cm]{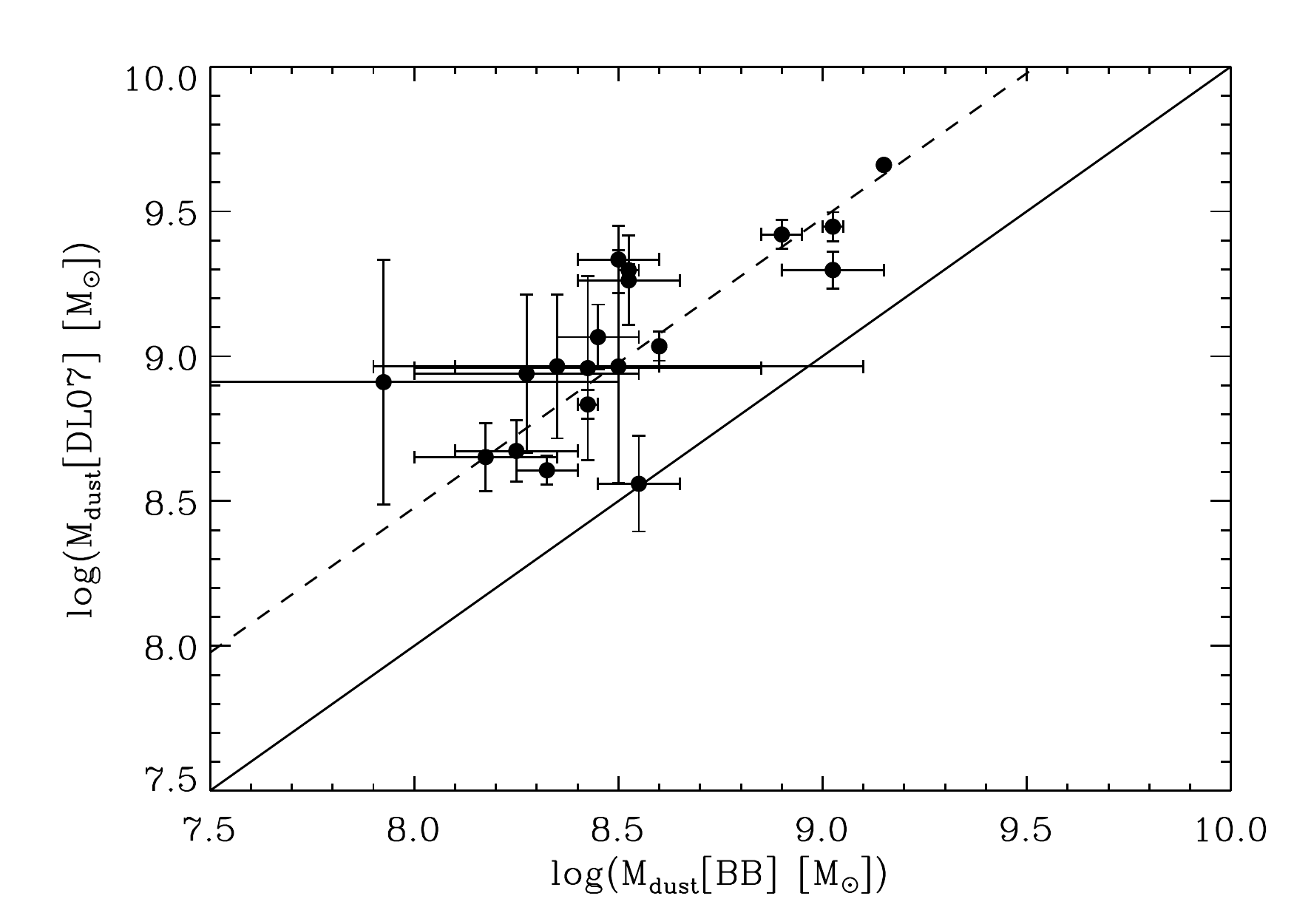}
\caption{Comparison of the dust masses estimated using the DL07 model with those estimated using a single temperature modified blackbody function.
The solid line corresponds to the one-to-one relation while the dashed line is offset by a factor 3.
We quote the geometric mean of $M_{{\rm dust}}$ across the region of parameter space where $\chi^2<\chi_{{\rm min}}^2+1$ and from which we define its 1$\sigma$ uncertainty.
\label{fig:mdust_compa}}
\end{figure}

\indent{
From the normalization of the modified blackbody function we can also compute a dust mass:\\}
\begin{equation}
M_{{\rm dust}}=\frac{S_{\nu}D_{{\rm L}}^2}{(1+z)\kappa(\nu_{0})B_{\nu}(\lambda_{{\rm rest}},T_{{\rm dust}})}\ \bigg( \frac{\nu_{0}}{\nu}\bigg)^{\beta},
\end{equation}
where $\kappa(\nu_{0})$ is the dust mass absorption cross section at the reference frequency $\nu_{0}$.
To allow direct comparison with the dust masses estimated using the DL07 model, we use the same dust mass absorption cross section at $\nu_{0}=c/100\,\mu$m, i.e., $\kappa(\nu_{0})=3.13$~m$^{2}\,$kg$^{-1}$ \citep{li_2001}.
Dust temperatures and dust masses inferred from this single modified blackbody function are given in Table \ref{tab: mdust}.
Here again we quote the geometric mean across the region of parameter space satisfying $\chi^2<\chi_{{\rm min}}^2+1$ and from which we define the 1$\sigma$ uncertainties.\\
\indent{
Fig. \ref{fig:mdust_compa} compares the dust masses obtained using a single modified blackbody function to those inferred using the DL07 model.
Estimates based on a single modified blackbody function are systematically lower by a factor $\thicksim3$.  This finding is in perfect agreement with that of \citet{dale_2012} based on local galaxies observed as part of the SINGS and KINGFISH surveys \citep[see also][]{magdis_2012}. These discrepancies arise from the fact that single temperature models do not take into account warmer dust emitting at shorter wavelength.  The behavior is also consistent with the analytic radiative transfer work of \citet{chakrabarti_2008}, who found that in high-redshift star-forming galaxies, dust masses inferred from isothermal models (equivalently single temperature blackbody approximation) are lower by a factor of $\thicksim\,$$2$ than those inferred using multi-$T$ components (see equation A8 of Chakrabarti and McKee \citeyear{chakrabarti_2008}).
In the rest of the paper we  work with dust masses inferred using the DL07 model as they account for the warmer dust emitting at shorter wavelength.
This choice is also motivated by the need to reduce any possible correlations between our dust temperature and dust mass estimates (see Section \ref{results} for more details). 
\\}
\subsection{Stellar masses and metallicities\label{subsec: stellar masses}}

In order to estimate the stellar masses of our galaxies we use the multi-wavelength catalogue built by the PEP consortium and presented in \citet{berta_2010,berta_2011}. It is a $z+K$-selected catalogue containing photometry in 16 bands from the \textit{Galex} far-UV to \textit{Spitzer} near-IR wavelengths. Stellar masses were calculated by fitting these multi-wavelength photometry to \citet{bruzual_2003} templates using FAST (Fitting and Assessment of Synthetic Templates; Kriek et al. \citeyear{kriek_2009}) and prescriptions presented by \citet{wuyts_2011b}.
We limit the \citet{bruzual_2003} templates to models with exponentially declining SFHs and a minimum $e$-folding time of 300 Myr.
This allows for a much better agreement between SFRs derived from optical-to-near-IR SED fits and those derived using mid/far-infrared observations.  
Full details are presented in \citet{wuyts_2011a} and \citet{wuyts_2011b}. 

\indent{
Metallicities are derived using the stellar-mass-metallicity relation at the redshift of the source (Genzel et al. \citeyear{genzel_2012}; see also Erb et al. \citeyear{erb_2006}; P. Buschkamp et al., in prep.; Shapley et al. \citeyear{shapley_2005}; Liu et al. \citeyear{liu_2008}):\\}
\begin{equation}
\label{eq1}
Z_{0}=2.18{\rm\,log}(M_{\ast})-0.0896 {\rm\,log}(M_{\ast})^2+a,
\end{equation}
where $a=-4.45$ if $1.0<z<1.5$ and $a=-4.51$ if $z>1.5$.   All metallicities are converted into the \citet{denicolo_2002} calibration system using the conversion function of \citet{kewley_2008}.
The Denicol\'o et al. calibration system provides the best agreement, at high stellar masses, between all different metallicity calibrators \citep{kewley_2008}.
Table \ref{tab: mdust} presents our inferred metallicities.
We note that metallicities estimated using the ``fundamental metallicity relation'' of \citet{mannucci_2010} extended to $z\thicksim2$ in \citet{genzel_2012} are consistent with those derived using equation \ref{eq1}.

\section{The ${\rm CO} \rightarrow H_2$ conversion factor}
\label{alphaco}
The total molecular gas mass is commonly derived from the measured CO line luminosity using a conversion factor \xco: 

\begin{equation}
M_{{\rm H_2}}=\alpha_{{\rm CO}} L^{\prime}_{{\rm CO(1-0)}},
\end{equation} 
where $L^{\prime}_{{\rm CO(1-0)}}$ is in (K \kms\ pc$^{2}$).  The exact value of \xco\ and its possible dependency on a range of parameters were and are currently being actively investigated, both with observations and simulations \citep[e.g.,][]{solomon_1987,dickman_1986,maloney_1988,israel_2000,bolatto_2008,leroy_2011,magdis_2011b,genzel_2012,glover_2011,narayanan_2011b,shetty_2011a,shetty_2011b,schruba_2012}.\\
\indent{
There is a general consensus that within the MW \xco$=4.34$ \msun (K \kms\ pc$^{2}$)$^{-1}$, with little spatial variations \citep[e.g.][]{strong_1996,dame_2001,abdo_2010}\footnote{We note that all the values of \xco\ quoted in this paper already include an upward factor of 1.36 to account for the presence of Helium coexisting with the molecular hydrogen.}. There is also ample evidence for at least two sets of circumstances where \xco\ departs significantly from the Galactic value:  low metallicity environments and the very dense central regions of merging systems.  In low metallicity conditions, the CO molecules are lacking the dust shielding they require to survive the ambient radiation field everywhere but in the cores of GMCs, while the H$_2$ molecules can survive in self-shielded envelopes around these cores \citep[e.g.][]{wolfire_2010}.  The result of this preferential photodissociation of the CO molecule are values of \xco\ larger by a factor of at least 5-10 (or even more) in local low-metallicity galaxies such as the SMC \citep{israel_1997,johansson_1998,leroy_2011,bolatto_2011,schruba_2012}.  On the other hand, in systems undergoing strong starbursts triggered by major mergers, such as nearby ULIRGs and high$-z$ SMGs, the value of \xco\  is reduced by a factor of 5 (Solomon et al. \citeyear{solomon_1997}; Downes \& Solomon \citeyear{downes_1998}; Tacconi et al. \citeyear{tacconi_2008}; but see Papadopoulos et al. \citeyear{papadopoulos_2012}).
\\}
\indent{
While these boundary conditions are generally accepted and routinely applied, the transitions between these different regimes are still being explored.  Here we combine the IRAM molecular gas data and the {\it Herschel} dust masses to estimate a conversion factor for each galaxy in our combined sample, and compare the values with different prescriptions for \xco\ variations. 
}
\subsection{The gas-to-dust ratio method}
Assuming that we can predict the gas-to-dust mass ratio \gdr\ in a galaxy, and having accurate molecular gas and dust masses, the conversion factor \xco\ can be estimated as:
\begin{equation}
\alpha_{{\rm CO}}=\frac{ \delta_{{\rm\scriptsize{GDR}}}(\mu_0) M_{{\rm dust}}}{L^{\prime}_{{\rm CO(1-0)}}}
\label{eq_xco}
\end{equation}
To estimate \gdr\ for each galaxy, we use the relation of \citet{leroy_2011} which relates it to metallicity, i.e., $\mu_0$:  
\begin{equation}
{\rm log}\,(\,\delta_{{\rm\scriptsize{GDR}}}(\mu_0)\,)=-0.85\times\mu_0+9.4. 
\label{eq_gdr}
\end{equation}
Recently, \citet{magdis_2011b} have shown that this ``dust-to-gas ratio method"  constrains \xco\ in two well-studied GOODS-N galaxies to values consistent with standard expectations.  However, we wish to point out three important assumptions required to extend this method from the local Universe to our high redshift galaxies:

\begin{enumerate}
\item Since the CO observations were done in large part with the PdBI in compact configuration, and since the {\it Herschel} PSF even at the shortest wavelengths has FWHM$\sim 6$\arcsec (i.e., corresponding to linear scales of $\thicksim\,$48$\,$kpc at $z$$\,\thicksim\,$$1$), we can only work with integrated dust and gas measurements.  Therefore, we must rely on the assumption that the total CO(2-1) line fluxes and the FIR fluxes we measure are emitted from the same physical regions.  In the local Universe, IR emission and CO measurements are seen to correlate well, with a tight relation between $L_{{\rm FIR}}$ and $L_{{\rm CO}}$ \citep[e.g.][]{sanders_1985,sanders_1991,gao_2004}, which can be interpreted as evidence for them being produced in the same region.  Furthermore, in local normal star-forming galaxies, the radial profiles of CO(2-1) line emission and of IR flux, as traced by 24$\,\mu$m emission, are strikingly similar \citep{leroy_2008}.  

\item Equation \ref{eq_xco} is built under the assumption that \mh $\gg$ \mhi, and therefore that $M_{{\rm gas}} \simeq$ \mh $= \delta_{{\rm\scriptsize{GDR}}}(\mu_0)\,M_{{\rm dust}}$.  While in nearby galaxies, on average \mhi $\sim 3$\mh, albeit with large galaxy-to-galaxy variations \citep{saintonge_2011a}, there are multiple reasons to suspect that the balance between atomic and molecular gas is significantly different in high redshift galaxies (see e.g., Obreschkow \& Rawlings \citeyear{Obreschkow_2009}).  First, the molecular gas mass fraction $f_{{\rm H_2}}=M_{{\rm H_2}}/(M_{\ast}+M_{{\rm H_2}})$, observed locally to be on average $6 \%$ in star-forming galaxies \citep{saintonge_2011a}, is found to be $30-50 \%$ at $z=1-2$ \citep[e.g.,][]{tacconi_2010,daddi_2010,geach_2011}.  It is therefore difficult to accommodate a significant atomic gas component within the total mass budget of these high-$z$ galaxies as inferred from dynamical analysis.  Observations of damped Lyman-$\alpha$ systems also suggest that the evolution of $\Omega_{\rm HI}$ between $z=0$ and $z=2$ is not as strong \citep[see e.g., figure 2 of ][]{Obreschkow_2009}.  Finally,  both models and observations indicate that above a critical surface density of $\sim 10$ \msun\ pc$^{-2}$ most of the gas is found in molecular form, with HI saturating at that threshold \citep[e.g.,][]{wong_2002,schaye_2004,blitz_2006,bigiel_2008}.  Since high-$z$ star-forming galaxies have high surface densities, we can therefore expect minimal contributions from HI over the regions of the disks where the CO and FIR emission are originating. 

\item We also work under the assumption that the local relation between \gdr\ and metallicity applies in the denser ISM of high-redshift galaxies.
In denser environments, the depletion of heavy elements from the gas phase is expected to increase \citep{jenkins_2009}, leading to lower \gdrm\ values.
However, in the denser environment of local-ULIRGs, only a mild decrease of \gdrm\ is observed, compared to normal star-forming galaxies  \citep{solomon_1997,wilson_2008,santini_2010,clements_2010}.  Thus, even if  high-redshift galaxies exhibit high surface densities, the effect on \gdrm\ should not exceed a factor 2.
The consequence of this effect is further discussed in Section \ref{results}.
Due to their higher surface densities, high redshift galaxies might also have higher dust emissivity \citep[see, e.g.,][]{Arce_1999,dutra_2003,cambresy_2005,michalowski_2010b} than local galaxies.
This would affect our dust mass estimates.
However, we show in Section \ref{results} than this effect does not strongly affect our main results.
\end{enumerate}
Based on these assumptions, we can combine the IRAM molecular gas data and the {\it Herschel} dust masses to estimate $\alpha_{CO}$ in each of the GOODS-N galaxies.

\subsection{Results \label{results}}
\begin{figure*}
\includegraphics[width=18.5cm]{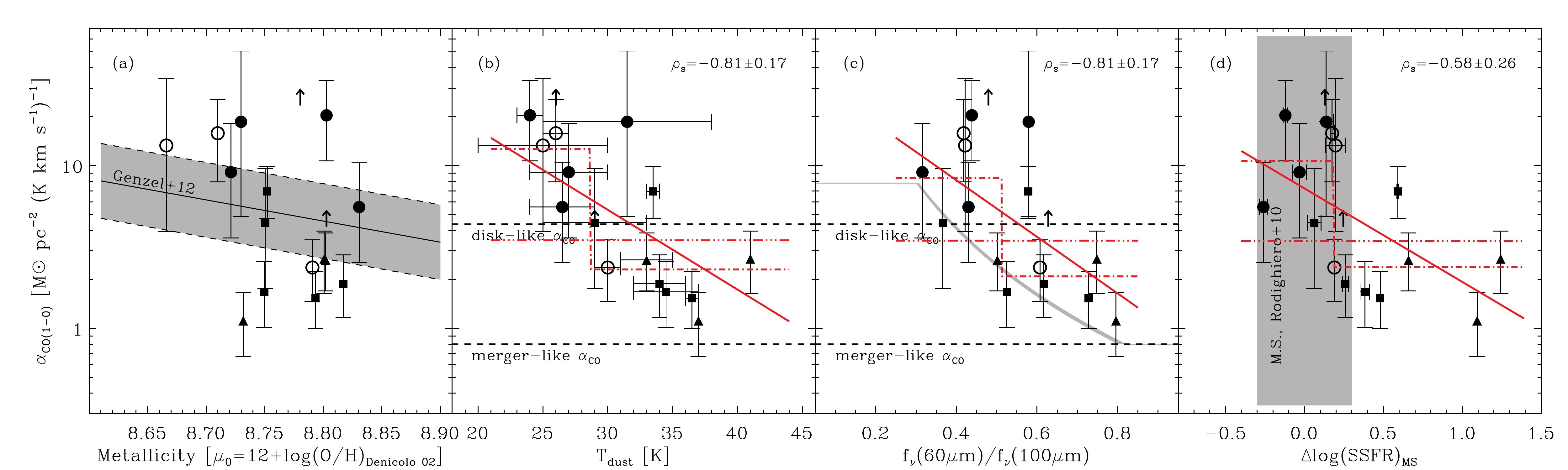}
\caption{
$\alpha_{{\rm CO(1-0)}}$ as function of metallicity (i.e., $\mu_{0}$), dust temperature derived from fitting a single modified blackbody function (i.e., $T_{{\rm dust}}$, see Section \ref{subsec: dust masses}), rest-frame far-infrared color (i.e., $f_{60}/f_{100}$) and distance with respect to the main sequence of star-formation (i.e., $\Delta$log$({\rm SSFR})_{MS}$).
Filled and opened circles represent our 4 secure and 3 tentative IRAM CO detections, respectively.
Our 2 IRAM CO upper limits are shown by arrows.
BzK galaxies and SMGs are shown by filled squares and triangles, respectively.
The shaded area in panel (a) represents the $\alpha_{{\rm CO(1-0)}}$-$\mu_{0}$ relation and its 1$\sigma$ uncertainty as found in \citet{genzel_2012}.
The shaded area in panel (d) represents the 1$\sigma$ scatter around the MS of \citet{rodighiero_2010b}.
The grey line in panel (c) represents the empirical $\alpha_{{\rm CO(1-0)}}$-$f_{60}/f_{100}$ relation observed in the local Universe by \citet{gracia_carpio_2011}.
Red lines in panel (b), (c) and (d) present fits to the data points using a constant (dashed-triple-dotted line), a step (dashed-dotted line) and a linear (solid) function.
Red lines are intentionally limited to the parameter space in which we have direct constraints.
\label{fig: alpha_co_predict}}
\end{figure*}

Figure \ref{fig: alpha_co_predict} shows the value of $\alpha_{{\rm CO}}$ as a function of metallicity ($\mu_{0}$), dust temperature ($T_{{\rm dust}}$), far-infrared color ($f_{60}/f_{100}$) and distance of a galaxy with respect to the MS of star-formation ($\Delta$log$({\rm SSFR})_{MS}$).
Our galaxies roughly follow the $\alpha_{{\rm CO}}$-$\mu_{0}$ relation found in \citet{genzel_2012}, although with a larger scatter ($\thicksim\,$0.5$\,$dex as opposed to the expected $\thicksim\,$$0.2\,$dex) and over a limited metallicity range which makes the slope uncertain.
We note that this larger scatter is expected because our sample contains normal SFGs and merger-driven starbursts as opposed to \citet{genzel_2012} who studied normal SFGs.
This scatter however correlates with $T_{{\rm dust}}$, $f_{60}/f_{100}$ and $\Delta$log$({\rm SSFR})_{MS}$.  
We note that \citet{gracia_carpio_2011} also proposed, in the local Universe, for a smooth evolution of $\alpha_{{\rm CO}}$ with $f_{60}/f_{100}$.
Their empirical relation is shown in the $\alpha_{{\rm CO}}$-$f_{60}/f_{100}$ panel of Fig. \ref{fig: alpha_co_predict}.
Despite some possible discrepancies in the absolute normalization of our two correlations, we find that the relative variation of $\alpha_{{\rm CO}}$ with $f_{60}/f_{100}$ found in our two studies are in excellent agreement.
Our $\alpha_{{\rm CO}}$-$\Delta$log$({\rm SSFR})_{MS}$ relation is also consistent with the results of \citet{magdis_2012} who find that galaxies situated off the MS exhibit lower $\alpha_{{\rm CO}}$ than galaxies on the MS of star-formation.

In order to characterize the correlation of $\alpha_{{\rm CO}}$ with the different quantities, we fit these relations with (i) a  constant, (ii) a step and (iii) a linear function.
To take into account errors both in X and Y, we use a Monte-Carlo approach.
We create $1\,000$ mock samples with the same number of sources as our original sample.
For each realization and for each galaxy, we attribute new values of $\alpha_{{\rm CO}}$, $T_{{\rm dust}}$, $f_{60}/f_{100}$ and $\Delta$log$({\rm SSFR})_{MS}$ selected into a Gaussian distribution centered at their original values and with a dispersion given by their errors.
We then fit each Monte-Carlo realization with our three different models (i.e., constant, step and linear functions).  Table \ref{tab: parameter} summarizes the results.  For each model, we quote the mean value and dispersion of each fitting parameter across the $1\,000$ Monte-Carlo realizations.  To assess the degree of correlation between \xco\ and $T_{{\rm dust}}$, $f_{60}/f_{100}$ and $\Delta$log$({\rm SSFR})_{MS}$, we also perform Spearman's test on each of the Monte-Carlo realizations, the results of which are also given in Table \ref{tab: parameter}.   The test reveals that the correlation of \xco\ with either $T_{{\rm dust}}$ or $f_{60}/f_{100}$ is stronger than with $\Delta$log$({\rm SSFR})_{MS}$. 

For the three relations shown in Figure \ref{fig: alpha_co_predict}b-d, the constant-\xco\ model always leads to higher reduced $\chi^{2}$ values than the step and linear functions, and can thus be rejected.   For the $\alpha_{{\rm CO}}$-$f_{60}/f_{100}$ correlation, the step and linear models are statistically undistinguishable, while in the case of the $\alpha_{{\rm CO}}$-$T_{{\rm dust}}$ and $\alpha_{{\rm CO}}$-$\Delta$log$({\rm SSFR})_{MS}$ correlations, the step function is statistically slightly better than the linear model.  However, there is a large overlap in the reduced $\chi^2$ distributions of the two models, preventing us to adopt or reject either.  We note that in the case of the step function,  the values of \xco\ below and above the break differ by a factor of 5, consistent with the relative difference between the Galactic and starburst \xco\ values measured in the local Universe \citep[see, e.g.,][]{solomon_1997}.  Although we cannot distinguish between a step and a linear model, we can unambiguously conclude that $\alpha_{{\rm CO}}$ strongly correlates with $T_{{\rm dust}}$ and $f_{60}/f_{100}$ and in a weaker way with $\Delta$log$({\rm SSFR})_{MS}$.  Before discussing the meaning of these correlations (Section \ref{sec: discussion}), we first discuss their robustness against systematic biases in the quantities involved in Eq. \ref{eq_xco}, i.e., \gdrm, $M_{{\rm dust}}$ and $\mu_0$.

\begin{table}
\scriptsize
\caption{\label{tab: parameter} Parameter values of the $\alpha_{{\rm CO}}$-$T_{{\rm dust}}$, $\alpha_{{\rm CO}}$-$f_{60}/f_{100}$ and $\alpha_{{\rm CO}}$-$\Delta$log$({\rm SSFR})_{MS}$ correlations}
\centering
\begin{tabular}{ c c c c c} 
\hline \hline
 \rule{0pt}{3ex}Model & & $\alpha_{{\rm CO}}$-$T_{{\rm dust}}$  & $\alpha_{{\rm CO}}$-$f_{60}/f_{100}$ & $\alpha_{{\rm CO}}$-$\Delta$log$({\rm SSFR})_{MS}$\\
\hline
 \rule{0pt}{3ex}Spearman's factor & & $-0.81\pm0.17$ & $-0.81\pm0.17$ & $-0.58\pm0.26$ \\
\hline
 \rule{0pt}{3ex}Constant$^{a}$ & $\chi^2_{{\rm red}}$ &  $5.2$ & $5.2$ & $6.2$\\
 & $a$ & $0.54\pm0.06$ & $0.54\pm0.06$ & $0.54\pm0.06$ \\
\hline
 \rule{0pt}{3ex}Step$^{b}$ & $\chi^2_{{\rm red}}$ & $3.1$ & $3.4$ & $3.4$ \\
 & $a$ & $1.11\pm0.18$ & $0.93\pm0.24$ & $1.03\pm0.15$ \\
 & $b$ & $0.37\pm0.09$ & $0.32\pm0.14$ & $0.38\pm0.08$ \\
 & $c$ & $28.7\pm2.7$ & $0.51\pm0.09$ & $0.18\pm0.09$ \\
 \hline
  \rule{0pt}{3ex}Linear$^{c}$ & $\chi^2_{{\rm red}}$ & $3.5$ & $3.6$ & $4.2$ \\
 & $a$ & $2.21\pm0.53$ & $1.61\pm0.29$ & $0.86\pm0.09$ \\
 & $b$ & $-0.049\pm0.017$ & $-1.74\pm0.49$ & $-0.57\pm0.14$\\
 \hline
\end{tabular}
\begin{list}{}{}
\item[\textbf{Notes.} ]
\item[$^{\mathrm{a}}$]  log($\alpha_{{\rm CO}}$)$\,=\,$$a$  
\item[$^{\mathrm{b}}$] log($\alpha_{{\rm CO}}$)$\,=\,$$a$ if $x$$\,<\,$$c$ and log($\alpha_{{\rm CO}}$)$\,=\,$$b$ if $x$$\,>\,$$c$, where $x$$\,\equiv\,$$T_{{\rm dust}}$, $f_{60}/f_{100}$ or $\Delta$log$({\rm SSFR})_{MS}$
\item[$^{\mathrm{a}}$]  log($\alpha_{{\rm CO}}$)$\,=\,$$a+b\times x$, where $x$$\,\equiv\,$$T_{{\rm dust}}$, $f_{60}/f_{100}$ or $\Delta$log$({\rm SSFR})_{MS}$
\end{list}

\end{table}

\noindent \textit{(i) Uncertainties on the \gdrm\ relation :}

We have assumed that the $z\sim1$ GOODS-N star-forming discs follow the local \gdrm\ relation derived by \citet{leroy_2011}. 
Instead, this relation might evolve with redshift, both in terms of normalization and slope.
Because $\alpha_{{\rm CO}}$ linearly scales with \gdrm\ (see Eq. \ref{eq_xco}), a different \gdrm\ relation at high-redshift would scale upward or downward the inferred values of $\alpha_{{\rm CO}}$.  Since all but two of our galaxies are situated in a narrow range of redshifts (i.e., $1.0<z<1.6$) and a narrow range of metallicities, a change in the normalization of the \gdrm\ relation with redshift would only translate into a re-normalization of our correlations, leaving their slopes mostly unchanged. 
Furthermore, since galaxies with cold and hot dust temperatures span similar metallicity ranges (Fig. \ref{fig: alpha_co_predict}; it would also be the case using metallicities inferred with the ``fundamental metallicity relation'' of Mannucci et al. \citeyear{mannucci_2010}, see point (iii)), a change of the slope of the \gdrm\ should not affect significantly our correlations.  For example, the correlation of $\alpha_{{\rm CO}}$ with $T_{{\rm dust}}$, $f_{60}/f_{100}$ and $\Delta$log$({\rm SSFR})_{MS}$ is recovered if using a constant value for the gas-to-dust ratio (150) rather than the metallicity dependent \gdrm. To produce the relations of Fig. \ref{fig: alpha_co_predict}, \gdrm\ would have to vary differentially for galaxies with cold and hot dust temperatures.
High dust temperatures are usually associated with the dense star-forming regions of major merger-driven starbursts \citep{sanders_2003,chapman_2003,magnelli_2012,hayward_2012}.  In these denser environments, the depletion of heavy elements from the gas phase is expected to increase \citep{jenkins_2009}, leading to lower \gdrm\ values.  Values of \gdrm\ in the dense environment of local ULIRGs are indeed found to be smaller than in normal star-forming galaxies \citep{solomon_1997,wilson_2008,santini_2010,clements_2010}, by a factor of at most  $\thicksim\,$2.  Taking into account this variation of \gdrm\ with density would however {\it strengthen} the correlation of $\alpha_{{\rm CO}}$ on e.g. $T_{{\rm dust}}$.   
Indeed, Eq. \ref{eq_xco} clearly shows that a lower value of \gdrm\ for galaxies with high densities and hot dust would result in an even smaller $\alpha_{{\rm CO}}$ compared to the normal star-forming galaxies with colder dust temperatures. We therefore conclude that it is unlikely that the uncertainties on the high-redshift \gdrm\ relation can artificially produce the observed correlation of $\alpha_{{\rm CO}}$ with $T_{{\rm dust}}$, $f_{60}/f_{100}$ and $\Delta$log$({\rm SSFR})_{MS}$.

\noindent \textit{(ii) Uncertainties on our dust mass estimates :}

We have followed the approach of \citet{leroy_2011} in following DL07 to infer dust masses. Consequently, any biases introduced by the DL07 method would affect both our dust mass estimates and their \gdrm\ prescription.
This consistency (also in term of rest-frame wavelength coverage, see Section \ref{subsec: dust masses}) protects $\alpha_{{\rm CO}}$ from systematic biases introduced by the methodology selected to calculate the dust masses. To erase our correlations, any dust mass bias would have to be specific to our galaxy sample, and correlate with their dust temperatures.
Compared to the objects studied in \citet{leroy_2011}, one specificity of our sample is that it also includes galaxies which are analogues of local-ULIRG, i.e., merger-driven starbursts like SMGs \citep[see, e.g.,][]{tacconi_2008,magnelli_2012}.   
Because these merger-driven starbursts have denser environments, they might have higher dust emissivity than normal star-forming galaxies \citep[see, e.g.,][]{Arce_1999,dutra_2003,cambresy_2005,michalowski_2010b}.  Therefore, using the DL07 method, we may have overestimated the dust masses of our merger-driven starbursts with hot dust temperatures.  Once again, because $\alpha_{{\rm CO}}$ linearly scales with $M_{{\rm dust}}$, correcting for this possible bias will not erase our correlations, but rather strengthen them.  In any case, we note that \citet{magnelli_2010,magnelli_2012} do not find any significant differences in the dust emissivity of high-redshift SMGs and lensed-SMGs.
\citet{magdis_2012} find that the absence of millimeter measurements might affect in a different way the dust mass estimates of normal MS galaxies and SMGs.
While for normal MS galaxies the dust masses estimated with or without millimeter measurements are fully consistent, for SMGs they find that without millimeter measurements dust masses seem to be overestimated by a factor of $\thicksim$$\,2$.
Among the three SMGs of our sample only one has a millimeter measurement.
In any case, because this effect would decrease the inferred $\alpha_{{\rm CO}}$ value of our SMGs, it would strengthen our $\alpha_{{\rm CO}}$-$T_{{\rm dust}}$ ($f_{60}/f_{100}$) correlation.
Finally, one might argue that $T_{{\rm dust}}$ and $M_{{\rm dust}}$ being both estimated from the same FIR measurements, they are artificially correlated.
For a given set of FIR measurements, higher $T_{{\rm dust}}$ would lead to lower values of $M_{{\rm dust}}$ and consequently to lower values of $\alpha_{{\rm CO}}$.
However, we remind the reader that here this effect is at least reduced by the fact that $T_{{\rm dust}}$ and $M_{{\rm dust}}$ are not directly derived from the same method, the former being derived using a single modified blackbody model and the latter being derived using the DL07 model.
We also note that the $\alpha_{{\rm CO}}$-$f_{60}/f_{100}$ and $\alpha_{{\rm CO}}$-$\Delta$log$({\rm SSFR})_{MS}$ correlations are not affected by this effect.
From all these analyses we conclude that uncertainties on our dust mass estimates could not explain the observed correlations of $\alpha_{{\rm CO}}$ with $T_{{\rm dust}}$, $f_{60}/f_{100}$ and $\Delta$log$({\rm SSFR})_{MS}$.

\noindent \textit{(iii) Uncertainties on our metallicity estimates :}

\citet{mannucci_2010} found that the metallicity of local star-forming galaxies does not only depend on their stellar masses but also on their SFRs.
Using metallicities inferred with the ``fundamental metallicity relation'' of \citet{mannucci_2010} and revised in \citet{genzel_2012} for $z$$\,\thicksim\,$$2$ star-forming galaxies, we find no significant change in our correlations.
As a second sanity check we assume a constant value for \gdrm\ (i.e., independent of the metallicity) and verify that correlations are still observed.
Finally we note that to explain the $\alpha_{{\rm CO}}$-$T_{{\rm dust}}$ correlation with a metallicity effect, all galaxies with cold and hot dust temperature should have low and high metallicities, respectively, and the metallicity range probed by these two galaxy populations should be of $\thicksim$$1\,$dex.
This is unlikely, given the small uncertainties on our stellar mass estimates, and the dispersion of the mass-metallicity relation.  

We therefore conclude that uncertainties on the \gdrm\ relation, on our measured dust masses or on the inferred metallicities, are unlikely to artificially produce the $\alpha_{{\rm CO}}$ correlations shown in \ref{fig: alpha_co_predict}.  The normalization and slope of the relations are however subject to some of these biases.   We find a median value for $\alpha_{{\rm CO}}$ of $2.5$ at the high dust temperature end ($T_{{\rm dust}}$$\,>\,$$30\,$K), compared to the value of 1.0 found in local major mergers \citep{solomon_1997,downes_1998}.   Likewise, at low dust temperatures ($T_{{\rm dust}}$$\,<\,$$30\,$K) typical of normal star-forming galaxies, the median value of $\alpha_{{\rm CO}}$ is $10$, while a value of 4.34 \xcounits\ is found in the Milky Way \citep{strong_1996,dame_2001,leroy_2011,abdo_2010}.  

Other than the sources of uncertainty listed above, we identify two other factors that may lead to the systematic overestimation of $\alpha_{{\rm CO}}$ by a factor of $\thicksim\,$2: 
(i) \citet{leroy_2011} argue that, at fixed metallicity, changes in the dust emissivity and \gdr\ between the atomic and molecular phases of the ISM may have biased upward, by a factor of $1.5\,$-$\,2$, their estimate of \gdrm.  Because the gas in high-redshift galaxies is mainly in a molecular phase, we might consider this bias to be specific to their study.  Correcting downward the \gdrm\ relation of \citet{leroy_2011} by a factor of $1.5-2$ would bring our values $\alpha_{{\rm CO}}$ in line with those in analogous $z=0$ systems. 
(ii) If indeed the atomic gas phase in high-redshift galaxies is negligible, qualitatively its inclusion in Eq. \ref{eq_xco} would systematically lead to lower values of $\alpha_{{\rm CO}}$.

\section{Discussion}
\label{sec: discussion}
Using PdBI CO(2-1) measurements and far-infrared observations from \textit{Herschel}, we estimate the value of $\alpha_{{\rm CO}}$ for a sample of 17 high-redshift galaxies, sampling a large region of the SFR-\mstar\ plane. We find that $\alpha_{{\rm CO}}$ correlates with dust temperatures (i.e., $T_{{\rm dust}}$) and far-infrared color (i.e., $f_{60}/f_{100}$).
A weaker correlation with the distance from the main sequence of star-formation (i.e., $\Delta$log$({\rm SSFR})_{MS}$) is also observed. 
None of these correlations could be explained by systematic biases in our study, although their exact parametrization still remains uncertain.   Due to these uncertainties, here we only discuss our correlations in terms of relative variations of $\alpha_{{\rm CO}}$ with dust temperature, far-infrared color, and distance of a galaxy from the MS.

\begin{figure*}
\center
\includegraphics[width=18.0cm]{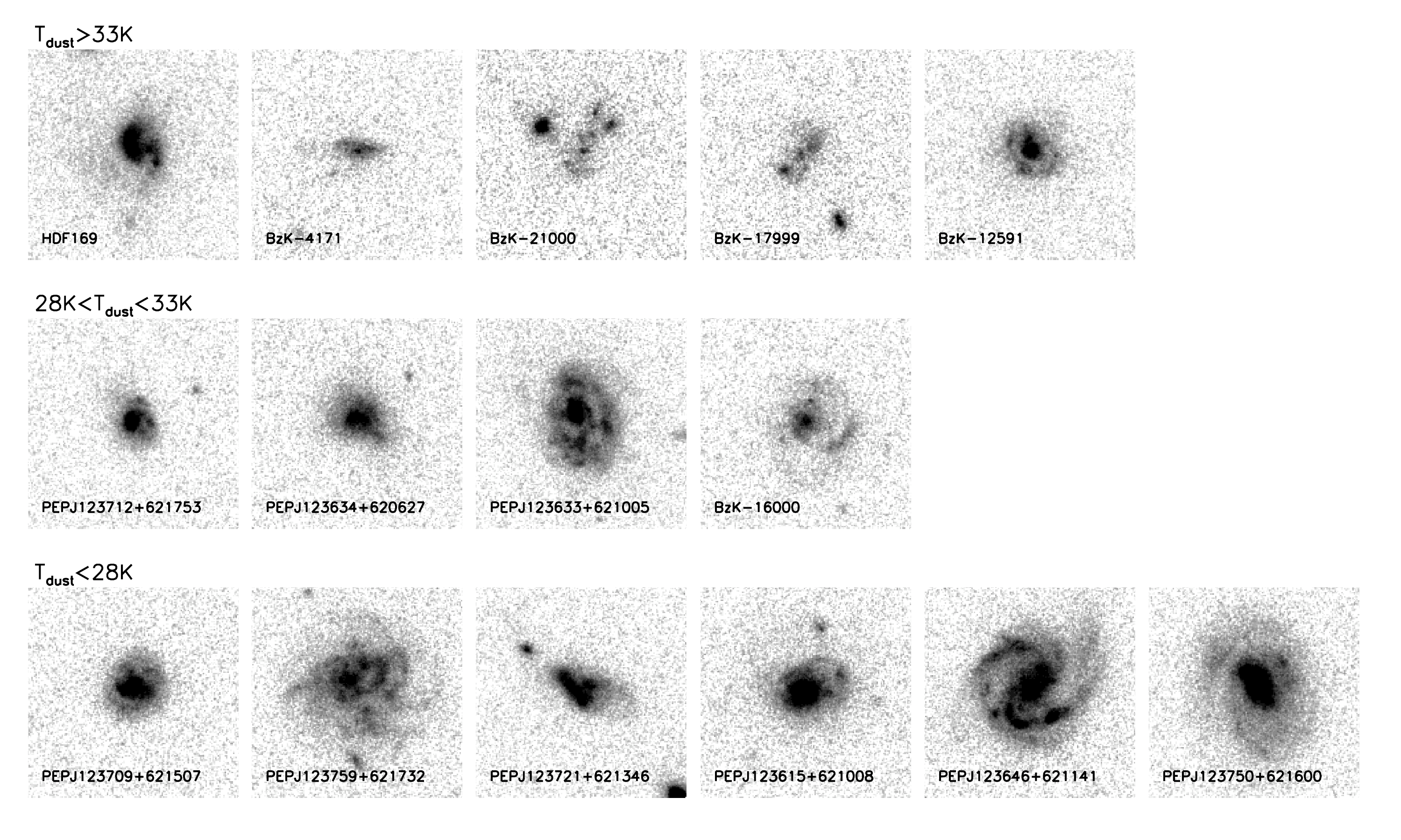}
\caption{ \label{Fig: stamps tdust} \textit{z}-band (i.e., F850LP) images from the \textit{Hubble} Space Telescope for galaxies in our sample with $1.0<z<1.6$ (i.e., all but two galaxies of our sample).
Each image is $40\,{\rm kpc}\times40\,{\rm kpc}$, and galaxies are ordered based on their dust temperatures. 
On average, the galaxies with $T_{{\rm dust}}<28\,$K have disk-like morphologies (i.e., PEPJ123709+621507, PEPJ123759+621732, PEPJ123615+621008, PEPJ123646+621141, PEPJ123750+621600) while most of the galaxies with $T_{{\rm dust}}>33\,$K are more compact and irregular/clumpy.
}
\end{figure*}
\subsection{The $\alpha_{{\rm CO}}$-$T_{{\rm dust}}$ and $\alpha_{{\rm CO}}$-$f_{60}/f_{100}$ correlations}

Although low CO rotational lines are almost always very optically thick \citep{dickman_1986,solomon_1987}, it has been shown in the MW and in nearby star-forming galaxies that CO line fluxes can be used to derive total molecular gas masses.  
The proportionality factor linking $L_{\rm CO}$ to $M_{H_2}$, $\alpha_{{\rm CO}}$, can be theoretically inferred using a ``virial'' approach.
At a given metallicity and for a self-gravitating (i.e. virialized) giant molecular cloud (GMC), $\alpha_{{\rm CO}}$ scales as:
\begin{equation}
\label{eq: alpha_co temperature}
\alpha_{{\rm CO}}\propto\frac{\langle n(H_2)\rangle^{1/2}}{T_b},
\end{equation}
where $\langle n(H_2)\rangle$ is the mean hydrogen density and $T_b$ is the CO intrinsic brightness temperature \citep[e.g.,][]{solomon_1987}.

Assuming that gas and dust are coupled in GMCs (i.e. $T_b\propto T_{{\rm dust}}$), equation \ref{eq: alpha_co temperature} may appear to suggest an anti-correlation between $\alpha_{{\rm CO}}$ and $T_{{\rm dust}}$.  However, observations of clouds in different environments of near-solar metallicity (in the MW and in other nearby normal star-forming galaxies) suggest that variations of $\langle n(H_2)\rangle$ are compensated by temperature changes, such that $\alpha_{{\rm CO}}$ remains constant, with variations of at most a factor of 2  \citep[e.g.][]{strong_1996,dame_2001,abdo_2010}.

In the local Universe, the only galaxies with near-solar metallicities yet significantly different values of $\alpha_{{\rm CO}}$ are the ULIRGs.  In these extreme merger-driven starbursts, the assumption that molecular gas clouds are in virial equilibrium breaks down, and measurements of  $\alpha_{{\rm CO}}$ suggest a value reduced by a factor of $\sim5$ compared to the Galactic value \citep[see, e.g.,][]{solomon_1997,downes_1998, scoville_1997,papadopoulos_2012}.  The scaling between $\alpha_{{\rm CO}}$, $\langle n(H_2)\rangle$, and $T_b$ however still holds, albeit with a lower proportionality factor \citep{solomon_1997,downes_1998,sakamoto_1999}, indicating that the low $\alpha_{{\rm CO}}$ values are mostly caused by a modification of the dynamics of their GMCs rather than by the increase in gas/dust temperature \citep[see, e.g.,][]{chapman_2003,clements_2010,wang_2011}.   These results suggest a step function of $\alpha_{{\rm CO}}$ as a function of $T_{{\rm dust}}$; normal SFGs dominated by virialized GMCs, which in general have cold dust temperatures, have $\alpha_{{\rm CO}}=4.36$ \xcounits, while  merger-driven starbursts, which always exhibit larger $T_{{\rm dust}}$ than normal SFGs, would have $\alpha_{{\rm CO}}\sim1$ \xcounits.
This interpretation qualitatively agrees with the step model fitted to our $\alpha_{{\rm CO}}$-$T_{{\rm dust}}$ and $\alpha_{{\rm CO}}$-$f_{60}/f_{100}$ correlations.
It also agrees quantitatively with it because in this model we find that galaxies with hot dust temperatures have \xco\ values five times lower than those with cold dust temperatures.
Finally, we also note that this interpretation might also agree with a simple visual inspection of the morphology of our galaxies (Fig.~\ref{Fig: stamps tdust}).
Galaxies with cold dust temperatures (i.e., $T_{{\rm dust}}<28\,$K) have on average a disk-like morphology characteristic of normal SFGs (i.e., PEPJ123709+621507, PEPJ123759+621732, PEPJ123615+621008, PEPJ123646+621141, PEPJ123750+621600), while galaxies with hot dust temperatures tend to be more compact and have disturbed morphologies (i.e., \textit{Bzk}-4171, \textit{Bzk}-21000, \textit{Bzk}-17999) consistent with being major-mergers.  Of course this conclusion has to be taken with caution because of low statistics of our sample and the difficulties encountered by any visual morphological classification.

Although satisfactory for our step model, this framework does not directly explain the possible smooth evolution of \xco\ with $T_{{\rm dust}}$ (or equivalently $f_{60}/f_{100}$), but can be extended to do so. High-redshift galaxies have larger gas fractions than in the local Universe \citep{tacconi_2010,daddi_2010,geach_2011}. 
Under these conditions, simple gravitational instabilities might locally create un-virialized GMCs \citep{Springel_2005,bournaud_2010}.
High-redshift galaxies may thus exist through a state where their ISM is a mixture of virialized and un-virialized GMCs (this mixture could either be observed at all physical scales, or else as a central region with un-virialized GMCs surrounded at larger scales by virialized GMCs).  The \textit{effective} $T_{{\rm dust}}$ and $\alpha_{{\rm CO}}$ derived from our unresolved observations would then correspond to a simple linear combination of these two ``types'' of star-forming regions, the cold virialized and hot un-virialized GMCs, creating the apparent smooth evolution of $\alpha_{{\rm CO}}$ with $T_{{\rm dust}}$ (or equivalently $f_{60}/f_{100}$).
The existence of two ``types'' of star-forming regions in interacting galaxies has already been found locally in the spatially resolved observations of Arp 158 (Boquien et al. \citeyear{boquien_2011}; see also Wei et al. \citeyear{wei_2012}).
While most star-forming regions of Arp 158 follow closely the Schmidt-Kennicutt relation found in spiral galaxies, its nuclear starburst and the tip of one of the tidal tails follow more closely the Schmidt-Kennicutt relation observed in merger-driven starbursts.
The existence of these two ``types'' of star-forming regions within the same galaxy seems to favor the interpretation where the smooth $\alpha_{{\rm CO}}$-$T_{{\rm dust}}$ relation is due to our spatially unresolved observations.

Using hydrodynamic simulations coupled with radiative transfer calculations, \citet{narayanan_2012} also find that \xco\ might evolve continuously from the Galactic to the merger-driven value.
They argue that at high redshifts gas rich galaxies with their large gravitational instabilities probe a larger dynamic range in ISM properties (gas/dust temperatures, density and velocity dispersion) than local galaxies, leading to a smooth evolution of \xco.  In particular, they find that mergers of different mass ratios lead to different conversion factors.

While we cannot formally discriminate between a step or a smooth evolution of \xco\ with the dust temperature or the  $f_{60}/f_{100}$ ratio, 
\textit{we can conclude that either quantity can be used to select an appropriate value of \xco; galaxies with hot dust temperatures unambiguously have lower \xco\ than galaxies with cold dust temperatures.}
The main advantage of $T_{{\rm dust}}$ for selecting the value of \xco\ is that it does not rely on the morphological classification of major-mergers/disky SFGs, which can be difficult at high $z$ because clumpy SFGs may be confused with major mergers.  
Moreover, even if the visual classifications of major-mergers/disky SFGs would be fully accurate, they would still be inefficient in distinguishing starbursting-mergers (i.e., with likely low \xco\ values) from late mergers with normal mode of star-formation (i.e., with likely high \xco values).
Indeed, we observe both locally and at high redshift that galaxies with unambiguous morphological signs of merging are not always associated with enhanced star formation activity \citep[e.g.][]{bushouse_1988,dasyra_2006,kartaltepe_2012}.
In that case, our $T_{{\rm dust}}$ criterion would be more indicative of the ISM conditions prevailing in these mergers.
We discuss further the implications of this result in section \ref{future}. 

 \subsection{The $\alpha_{{\rm CO}}$-$\Delta$log$({\rm SSFR})_{MS}$ correlation}

Currently, it is commonly interpreted that galaxies situated on the main sequence of the SFR-$M_{\ast}$ are consistent with a secular mode of star formation, while star-forming galaxies located far above the MS are powered by strong starbursts with short duty-cycles, likely triggered by major mergers \citep[][Magnelli et al., in prep.]{elbaz_2011,nordon_2012,rodighiero_2011,wuyts_2011b,magnelli_2012}.   In that view, a correlation between $\alpha_{{\rm CO}}$ and $\Delta$log$({\rm SSFR})_{MS}$ can therefore be expected: on-MS galaxies should have high $\alpha_{{\rm CO}}$ values while above-MS galaxies should have low $\alpha_{{\rm CO}}$ values as those seen in local ULIRGs.
Such a correlation is indeed observed in the right panel of Fig.~\ref{fig: alpha_co_predict}, though with a lower Spearman's correlation factor than those found for $T_{{\rm dust}}$ or $f_{60}/f_{100}$.  
However, we note that this lower Spearman's correlation factor might simply reflect the larger uncertainties in deriving $\Delta$log$({\rm SSFR})_{MS}$ than $T_{{\rm dust}}$ or $f_{60}/f_{100}$ (i.e., uncertainties on $M_{\ast}$, SFR and on the exact localization of the MS).  

Although we cannot say whether $\alpha_{{\rm CO}}$ depends on $\Delta$log$({\rm SSFR})_{MS}$ as a smooth or a step function, we can unambiguously conclude that galaxies situated far above the MS have lower value of $\alpha_{{\rm CO}}$ than galaxies situated on the MS.  Using $\Delta$log$({\rm SSFR})_{MS}$ as a proxy of the $\alpha_{{\rm CO}}$ value will be of great importance in the near future especially where deep PACS-SPIRE \textit{Herschel} observations are missing to derive $T_{{\rm dust}}$ and $f_{60}/f_{100}$.

The observed $\alpha_{{\rm CO}}$-$\Delta$log$({\rm SSFR})_{MS}$ correlation also directly confirms the current interpretation that galaxies on the main sequence are in a secular/steadier mode of star-formation, while off-MS galaxies are in a starbursting mode likely triggered by major mergers.  However, \citet{narayanan_2012} argue that high-$z$ galaxies, having high gas surface densities and strong gravitational instabilities, could have low $\alpha_{{\rm CO}}$ values without requiring a major-merger.   In any case, our observations allow us to conclude that  \textit{the different mechanisms having the power to raise a galaxy above the main sequence must also affect the mode of star-formation and the physical conditions prevailing in the star-forming regions}.  This interpretation is also supported by the observation that galaxies on and above the star formation MS have significantly different PAH abundances \citep[][]{nordon_2012} and star-formation rate surface densities \citep{wuyts_2011b}, other tracers of physical conditions in the ISM.

 \subsection{Implications for future work}
 \label{future}
  
Our observations confirm that at high redshift, just like in the local Universe, the value of $\alpha_{{\rm CO}}$ varies between galaxies with different star formation properties, corroborating results based on dynamical mass estimates in SMGs \citep{tacconi_2008}.  Applying a Galactic $\alpha_{{\rm CO}}$ to starbursting galaxies would not only lead to unrealistically high gas mass fractions \citep{genzel_2010}, but also to high \gdrm\ values, inconsistent with local observations and chemical evolution models.
  
More importantly, our results provide an empiral way to select the appropriate value of $\alpha_{{\rm CO}}$, using $T_{{\rm dust}}$ or $f_{60}/f_{100}$.  These indicators have the great advantage of being less subjective than visual morphological classifications of mergers/SFGs, and more indicative of ISM conditions than a fixed $L_{{\rm IR}}$ criterion.  In the absence of far-infrared measurements to derive dust properties, the offset of a galaxy from the star formation main sequence, $\Delta$log$({\rm SSFR})_{MS}$, is also a valid indicator. 

The galaxies studied here all have near-solar metallicities, as estimated using the mass-metallicity relation.  However, both observations and models show that $\alpha_{{\rm CO}}$ increases in low metallicity environments due to the photodissociation of CO, which lacks the power of H$_2$ to self-shield against radiation even when dust shielding becomes deficient \citep[e.g.][]{genzel_2012,glover_2011,narayanan_2011b,schruba_2012}.  The presence of an $\alpha_{{\rm CO}}$-$T_{{\rm dust}}$ correlation has yet to be tested at low metallicities, where the main driver of $\alpha_{{\rm CO}}$ is the availability or not of dust shielding.  One could expect $\alpha_{{\rm CO}}$ to vary along a plane defined by $T_{{\rm dust}}$ and metallicity, as suggested by the simulations of  \citet{narayanan_2012}, where a correlation between $\alpha_{{\rm CO}}$, gas surface density and merger mass ratio is observed even in low metallicity environments. 

The choice of $\alpha_{{\rm CO}}$ also has implications for the Schmidt-Kennicutt star formation relation, where SFGs and starbursting major-mergers form two distinct sequences \citep[][but see Ivison et al. \citeyear{ivison_2011}]{genzel_2010,daddi_2010}.  
If \xco\ evolves with the dust temperature (or alternatively $f_{60}/f_{100}$ or $\Delta$log$({\rm SSFR})_{MS}$) as a step function, it will not modify this observation. Adopting a value of $\alpha_{{\rm CO}}$ that varies smoothly with e.g. $T_{{\rm dust}}$ would not significantly change those results.
However, it would lead to a smoother transition where starbursting mergers with lower mass ratios or gas-rich disks with extreme gravitational instabilities would lie between these two sequences.

\section{Summary}
\label{summary}

We have combined deep {\it Herschel} PACS/SPIRE imaging and IRAM PdBI CO(2-1) measurements for a sample of 17 galaxies at $z>1.0$, selected out of the GOODS-N field to span a broad region of the SFR-\mstar\ plane.  The far-infrared data were used to derive accurate dust masses from the model of \citet{draine_2007}, while dust temperatures are recovered from the fitting of modified blackbody functions of fixed dust emissivity $\beta=1.5$.  

Using the prescription of \citet{leroy_2011} for the metallicity-dependence of the gas-to-dust ratio, the measured dust masses and the CO luminosities, we inferred the value of \xco\ in each of the galaxies.   The conversion factor is found to vary systematically with dust temperature and far-infrared color, two tracers of the physical conditions in star-forming regions.  We also found a weaker correlation between \xco\ and $\Delta$log$({\rm SSFR})_{MS}$, the distance from the sequence traced by star-forming galaxies in the SFR-\mstar plane.  We have investigated in detail possible systematic biases in the analysis, but find none that can convincingly explain these correlations. 

The results are consistent with previous observations indicating that in normal star-forming galaxies of near-solar metallicities \xco\ is mostly constant as long as star formation takes place in well virialized GMCs, while in major merger-driven starbursts where gas is compressed and dust temperatures are higher, the value of \xco\ is decreased by a factor of $\sim5$.
However, our measurements of \xco\ are consistent not only with a step function, but also with being a smooth function of $T_{\rm dust}$.
This smooth evolution may come from the fact that we are working with integrated measurements tracing the average conditions in the galaxies.
Either merging events with low mass ratios, or gravitational instabilities within the very gas-rich discs could lead to a situation where, within a single galaxy, some star formation is taking place within normal GMCs while some stars are also formed in dense, hot regions.  The integrated values of \xco\ and $T_{\rm dust}$ would then be expected to vary anywhere between the Galactic and the ``starburst" values, producing the smooth relation we observe. 
Alternatively, conditions may be changing at the scale of the star forming regions such that the \xco-$T_{\rm dust}$ relation would be recovered by resolved observations at the sub-kiloparsec scale.

Regardless of the exact functional form of the dependence of \xco\ on $T_{\rm dust}$, our study clearly shows that dust temperature (or equivalently the $f_{60}/f_{100}$ ratio) can be used to assign a single conversion factor value to any galaxy.  
Such a criterion is valuable, as it traces the physical conditions of the ISM more reliably than a simple SFR or $L_{{\rm IR}}$ criterion.  
It is also valuable because it is less subjective than methods requiring challenging visual classifications of major mergers which furthermore cannot easily distinguish between mergers in a starbursting phase from those in a normal phase of star-formation.
In the absence of far-infrared measurements to derive dust properties, the distance of a galaxy from the star formation main sequence can be used to make a choice between a Galactic conversion factor (\xco$=4.36$ \xcounits, including the contribution of Helium) and a ``starburst" value (\xco$=1.0$ \xcounits).  While most galaxies located above the main sequence are major mergers, other mechanisms may generate bursts of star formation powerful enough to displace the galaxies in the SFR-\mstar\ plane.  Our results however indicate that any such mechanism does so by modifying the effective mode of star formation and the physical conditions prevailing in the sites of star formation.

\acknowledgement{
We thank the staff of the IRAM Observatory in Grenoble for their support in preparing and conducting the observations, and in particular Yannick Libert and Jan-Martin Winters for assistance with data reduction. 
PACS has been developed by a consortium of institutes led by MPE (Germany) and including UVIE (Austria); KU Leuven, CSL, IMEC (Belgium); CEA, LAM (France); MPIA (Germany); INAF-IFSI/OAA/OAP/OAT, LENS, SISSA (Italy); IAC (Spain). This development has been supported by the funding agencies BMVIT (Austria), ESA-PRODEX (Belgium), CEA/CNES (France), DLR (Germany), ASI/INAF (Italy), and CICYT/MCYT (Spain).
SPIRE has been developed by a consortium of institutes led by Cardiff University (UK) and including University of Lethbridge (Canada), NAOC (China), CEA, LAM (France), IFSI, University of Padua (Italy), IAC (Spain), Stockholm Observatory (Sweden), Imperial College London, RAL, UCL-MSSL, UKATC, University of Sussex (UK), Caltech, JPL, NHSC, University of Colorado (USA). This development has been supported by national funding agencies: CSA (Canada); NAOC (China); CEA, CNES, CNRS (France); ASI (Italy); MCINN (Spain); SNSB (Sweden); STFC, UKSA (UK) and NASA (USA).}
\bibliographystyle{aa}

\end{document}